\documentclass[10pt,aps,preprintnumbers,prd,noshowpacs,nofootinbib,noshowkeys,floatfix,superscriptaddress]{revtex4}
\usepackage[dvips]{graphics,graphicx}
\usepackage[colorlinks=true,linktocpage=true,linkcolor=blue,citecolor=blue]{hyperref}
\usepackage[usenames,dvipsnames]{color}
\usepackage{amsmath, amssymb,oldgerm}
\usepackage{multirow}
\usepackage{longtable}
\usepackage{color}
\usepackage[normalem]{ulem}  
\usepackage{braket}
\usepackage{slashed}
\usepackage[mathscr]{eucal}

\newcommand{\Tr}{\text{Tr}}
\newcommand{\n}{\nonumber}

\newcommand{\s}{\mathfrak{s}}

\newcommand{\F}{\mathcal{F}}
\newcommand{\Pc}{\mathcal{P}}
\newcommand{\V}{\mathcal{V}}
\newcommand{\A}{\mathcal{A}}
\newcommand{\mC}{\mathfrak{C}}
\newcommand{\C}{\mathcal{C}}
\newcommand{\Sc}{\mathcal{S}}
\newcommand{\ms}{\mathfrak{s}}
\newcommand{\f}{\mathfrak{f}}
\newcommand{\D}{\mathcal{D}}

\newcommand{\beq}{\begin{equation}}
\newcommand{\eeq}{\end{equation}}
 \newcommand{\im}{\text{Im}\, }
 \newcommand{\re}{\text{Re}\, }

\newcommand{\free}{_{HW,\text{free}}}

\newcommand{\psbar}{\bar{\psi}}
\newcommand{\ps}{\psi}
\newcommand{\Sp}{\hat{S}}
\newcommand{\T}{\hat{T}}
\newcommand{\olra}{\overleftrightarrow}
\newcommand{\eqs}{Eqs.~}
\newcommand{\eq}{Eq.~}

\newcommand{\rrr}{\textcolor{red}}
\newcommand{\ePi}{\Pi}
\newcommand{\bA}{\mathbb{A}}
\newcommand{\Dd}{\nabla}

\newcommand{\avg}[1]{\left\langle:#1:\right\rangle}

\renewcommand{\d}{d}
\newcommand{\J}{\hat{{J}}}
\renewcommand{\S}{\hat{S}}
\newcommand{\odouble}{\olra}

\newcommand{\WD}[1]{W_{D,#1}}
\newcommand{\WP}[1]{W_{P,#1}}
\newcommand{\TD}[1]{T_{D,#1}}
\newcommand{\TP}[1]{T_{P,#1}}
\newcommand{\SD}[1]{S_{D,#1}}
\newcommand{\SP}[1]{S_{P,#1}}

\begin{document}

\title{Pseudo-gauges and relativistic spin hydrodynamics for interacting Dirac and Proca fields}

\author{Nora Weickgenannt}

\affiliation{Institute for Theoretical Physics, Goethe University,
Max-von-Laue-Str.\ 1, D-60438 Frankfurt am Main, Germany}

\author{David Wagner}

\affiliation{Institute for Theoretical Physics, Goethe University,
Max-von-Laue-Str.\ 1, D-60438 Frankfurt am Main, Germany}

\author{Enrico Speranza}

\affiliation{Illinois Center for Advanced Studies of the Universe and Department of Physics, University of Illinois at Urbana-Champaign, Urbana, IL 61801, USA}

\begin{abstract}
We present the explicit expressions of different pseudo-gauge transformations for Dirac and Proca fields considering a general interaction term. The particular case of the interaction of Dirac and Proca fields with a background electromagnetic field is also studied. Starting from the quantum kinetic theory with collisions derived from the Wigner-function formalism for massive spin-1/2 and spin-1 particles, we establish a connection between different pseudo-gauges and relativistic spin hydrodynamics.  The physical implications of the various decompositions of orbital and spin angular momentum are discussed. 
\end{abstract}




\maketitle

\section{Introduction}

The derivation of relativistic spin hydrodynamics, i.e., the theory of relativistic hydrodynamics when spin degrees of freedom are dynamical variables, has recently been the subject of intense research~\cite{Florkowski:2017ruc,Florkowski:2017dyn,Florkowski:2018myy,Florkowski:2018fap,Weickgenannt:2019dks,Bhadury:2020puc,Weickgenannt:2020aaf,Shi:2020htn,Speranza:2020ilk,Bhadury:2020cop,Singh:2020rht,Bhadury:2021oat,Peng:2021ago,Weickgenannt:2021cuo,Sheng:2021kfc,Sheng:2022ssd,Hu:2021pwh,Hu:2022lpi,Singh:2022ltu,Weickgenannt:2022zxs,Das:2022azr,Montenegro:2017rbu,Montenegro:2018bcf,Montenegro:2020paq,Gallegos:2021bzp,Gallegos:2022jow,Hattori:2019lfp,Fukushima:2020ucl,Li:2020eon,She:2021lhe,Wang:2021ngp,Wang:2021wqq,Daher:2022xon,Gallegos:2020otk,Garbiso:2020puw,Cartwright:2021qpp,Hongo:2021ona}. Such effort is mainly motivated by the phenomenology of noncentral heavy-ion collisions, where the vorticity of the hot and dense matter induces hadron spin polarization of the final state~\cite{Liang:2004ph,Voloshin:2004ha,Betz:2007kg,Becattini:2007sr}. 
Polarization phenomena for spin-1/2 particles have been observed in the case of Lambda hyperons in Refs.~\cite{STAR:2017ckg,Adam:2018ivw,STAR:2019erd}. The polarization of $\Lambda$-hyperons along the global angular momentum, i.e. the global polarization, was found to be in good agreement with hydrodynamic models assuming local thermodynamic equilibrium 
\cite{Becattini:2007sr,Becattini:2013vja,Becattini:2013fla,Becattini:2015ska,Becattini:2016gvu,Karpenko:2016jyx,Pang:2016igs,Xie:2017upb}. However, the same models cannot describe the momentum dependence of the polarization along the beam direction, i.e., the longitudinal polarization~\cite{Becattini:2017gcx}. This mismatch between experimental data and theoretical calculations is often referred to as ``polarization sign problem" and triggered many important developments, see, e.g.,~\cite{Becattini:2017gcx,Becattini:2020ngo,Florkowski:2019qdp,Florkowski:2019voj,Zhang:2019xya,Becattini:2019ntv,Xia:2019fjf,Wu:2019eyi,Sun:2018bjl,Liu:2019krs,Florkowski:2021wvk,Yi:2021ryh,Florkowski:2021xvy}. Recently, promising progress towards a restoration of the agreement between theory and experiments has been made in Refs.~\cite{Liu:2021uhn,Fu:2021pok,Becattini:2021suc,Becattini:2021iol,Liu:2021nyg}. Nevertheless, the polarization sign problem remains an open question. 
Furthermore, measurements of polarization observables have been also carried out for vector particles. In particular, the global spin alignment has been measured for $\phi$ and $K^{\star 0}$ mesons
~\cite{ALICE:2019aid,Mohanty:2021vbt}. Interestingly, the experimental results for the magnitude of the spin alignment turns out to be much larger compared to the theoretical predictions based on the assumption of local equilibrium of spin degrees of freedom~\cite{Liang:2004xn,Yang:2017sdk,Sheng:2019kmk,Sheng:2020ghv,Xia:2020tyd,Muller:2021hpe}. 

In order to resolve the disagreements between theory and experiments, it has been proposed to consider out-of-equilibrium spin effects in kinetic theory and include spin degrees of freedom as new dynamical variables in the hydrodynamic description of the hot and dense matter. In relativistic spin hydrodynamics, together with the conservation of the energy-momentum tensor $T^{\mu\nu}$,
one also solves the conservation of the total angular momentum tensor
\begin{equation}
J^{\lambda, \mu \nu} \equiv
x^\mu T^{\lambda \nu} - x^{\nu}T^{\lambda \mu} + \hbar S^{\lambda, \mu \nu},
\end{equation}
where $S^{\lambda, \mu \nu}$ is the so-called spin tensor. The macroscopic hydrodynamic quantities are given by the expectation values of the quantum operators over some (not necessarily equilibrium) state, i.e., $T^{\mu\nu}=\langle:\hat{T}^{\mu\nu}:\rangle$ and $S^{\lambda, \mu \nu}=\langle:\hat{S}^{\lambda, \mu \nu}:\rangle$, where the colon denotes normal ordering. Thus, the equations of motion for relativistic spin hydrodynamics read
\begin{subequations}\label{conslaws}
 \begin{eqnarray}
  \partial_\mu T^{\mu\nu}&=&0\; ,\\
 \hbar\, \partial_\lambda S^{\lambda,\mu\nu}&=& T^{\nu\mu}-T^{\mu\nu}  \;.
 \end{eqnarray}
 \end{subequations}
Over the past few years, such a theory has been developed from many different perspectives: kinetic theory~\cite{Florkowski:2017ruc,Florkowski:2017dyn,Florkowski:2018myy,Florkowski:2018fap,Weickgenannt:2019dks,Bhadury:2020puc,Weickgenannt:2020aaf,Shi:2020htn,Speranza:2020ilk,Bhadury:2020cop,Singh:2020rht,Bhadury:2021oat,Peng:2021ago,Weickgenannt:2021cuo,Sheng:2021kfc,Sheng:2022ssd,Hu:2021pwh,Hu:2022lpi,Singh:2022ltu,Weickgenannt:2022zxs,Das:2022azr}, 
an effective action~\cite{Montenegro:2017rbu,Montenegro:2018bcf,Montenegro:2020paq,Gallegos:2021bzp,Gallegos:2022jow}, 
an entropy-current analysis~\cite{Hattori:2019lfp,Fukushima:2020ucl,Li:2020eon,She:2021lhe,Wang:2021ngp,Wang:2021wqq,Daher:2022xon},
holographic duality~\cite{Gallegos:2020otk,Garbiso:2020puw,Cartwright:2021qpp}, and linear-response theory 
\cite{Montenegro:2020paq,Hongo:2021ona}. 
An important issue concerning the relativistic decomposition of the total angular momentum into an orbital and spin part is that the definition of the energy-momentum and spin tensors is not unique. In fact, one can perform a so-called pseudo-gauge transformation which is a redefinition of the densities of the form~\cite{Hehl:1976vr}
\begin{subequations}\label{PGT}
\begin{eqnarray}
  \T_\text{pgt}^{\mu\nu}&=& \T^{\mu\nu}+\frac{\hbar}{2}\partial_{\lambda}(\hat{\Phi}^{\lambda,\mu\nu}+\hat{\Phi}^{\nu,\mu\lambda}+\hat{\Phi}^{\mu,\nu\lambda})\; , \\
  \Sp_\text{pgt}^{\lambda,\mu\nu}&=& \Sp^{\lambda,\mu\nu}-\hat{\Phi}^{\lambda,\mu\nu}+\hbar\, \partial_\rho \hat{Z}^{\mu\nu\lambda\rho} \;,
 \end{eqnarray}
 \end{subequations}
 where $\hat{\Phi}^{\lambda,\mu\nu}$ and $\hat{Z}^{\mu\nu\lambda\rho}$ are arbitrary differentiable tensors such that $\hat{\Phi}^{\lambda,\mu\nu}=-\hat{\Phi}^{\lambda,\nu\mu}$ and $\hat{Z}^{\mu\nu,\lambda\rho}=-\hat{Z}^{\nu\mu,\lambda\rho}=-\hat{Z}^{\mu\nu,\rho\lambda}$. For convenience, in this work $\T^{\mu\nu}_\text{pgt}$ and $\Sp^{\lambda,\mu\nu}_\text{pgt}$ will always be constructed starting from the canonical tensors. The pseudo-gauge transformations have the properties to leave invariant the form of  \eqs\eqref{conslaws}, the global energy and momentum $\hat{P}^\mu$, and the global total angular momentum $\hat{J}^{\mu\nu}$ defined as
\begin{subequations}
 \begin{align}
     \hat{P}^\mu&\equiv \int d\Sigma_\lambda \hat{T}^{\lambda\mu}\; ,\\
     \hat{J}^{\mu\nu}&\equiv \int d\Sigma_\lambda \hat{J}^{\lambda,\mu\nu}\; ,
 \end{align}
 \end{subequations}
where $d\Sigma_\lambda$ denotes the integration over a space-like hypersurface.
Note that the global spin defined as
\begin{equation}
\hat{S}^{\mu\nu}\equiv \int d\Sigma_\lambda \hat{S}^{\lambda,\mu\nu}
\end{equation}
transforms as a rank-2 tensor if and only if the antisymmetric part of the energy-momentum tensor vanishes and the spin tensor is conserved \cite{Speranza:2020ilk}.
Recently, different choices of pseudo-gauges and their possible physical implications have been discussed in different contexts~\cite{Becattini:2011ev,Becattini:2012pp,Becattini:2018duy,Speranza:2020ilk,Fukushima:2020ucl,Li:2020eon,Buzzegoli:2021wlg,Das:2021aar,Daher:2022xon,Fukushima:2020qta,Leader:2013jra}. However, this topic still remains highly debated. 
While for free spin-1/2 particles pseudo-gauge transformations have been discussed in depth in the literature, see, e.g., Ref.~\cite{Speranza:2020ilk}, only little work has been devoted to studying pseudo-gauges for spin-1 particles and interacting spin-1/2 or spin-1 particles. In this paper, we aim at filling this gap.

In previous works~\cite{Weickgenannt:2020aaf,Weickgenannt:2021cuo,Weickgenannt:2022zxs}, relativistic dissipative spin hydrodynamics was derived from quantum kinetic theory for massive spin-1/2 particles with nonlocal collisions in the so-called Hilgevoord-Wouthuysen pseudo-gauge. In this paper, we provide a detailed derivation of the various sets of tensors (including the Hilgevoord-Wouthuysen ones used in~\cite{Weickgenannt:2020aaf,Weickgenannt:2021cuo,Weickgenannt:2022zxs}) by generalizing the pseudo-gauge transformations of free Dirac fields to the case of nonlocal interactions. Furthermore, we present for the first time the pseudo-gauge transformations for Proca fields, considering both the free and the interacting case. We find a form of the spin tensor analogous to the Hilgevoord-Wouthuysen one for spin-1/2 particles, which is conserved for free fields, but not in the presence of nonlocal collisions. Finally, we discuss pseudo-gauge transformations in the presence of electromagnetic fields for both massive spin-1/2 and spin-1 particles, obtaining a gauge-invariant splitting of the total angular-momentum tensor. This angular-momentum decomposition is such that the spin tensor is not conserved, but follows equations of motion analogous to the classical spin precession in electromagnetic fields~\cite{Bargmann:1959gz,Bailey:1975fe}. 

This paper is organized as follows. In Section \ref{intdirsec}, we briefly review the quantum kinetic theory for Dirac particles \cite{Weickgenannt:2020aaf,Speranza:2020ilk} and perform the pseudo-gauge transformations for Dirac fields interacting through a nonlocal collision term. In Section \ref{freeprocsec}, we introduce the energy-momentum and spin tensors for free Proca fields in various pseudo-gauges. As a next step, we generalize these pseudo-gauge transformations to the interacting case in Section \ref{intprocsec}. In Section \ref{eomsec}, we provide the equations of motion for relativistic spin hydrodynamics in the Hilgevoord-Wouthuysen pseudo-gauge. Such equations of motion are formally identical for spin-1 and spin-1/2 fields. Finally, in Section \ref{emfsec}, we discuss the particular case of the pseudo-gauge transformations in the presence of a background electromagnetic field.

We use the following notation and conventions: $a\cdot b\equiv a^\mu b_\mu$,
$a_{[\mu}b_{\nu]}\equiv a_\mu b_\nu-a_\nu b_\mu$, $a_{(\mu}b_{\nu)}\equiv a_\mu b_\nu+a_\nu b_\mu$, $g_{\mu \nu} = \mathrm{diag}(+,-,-,-)$,
$\epsilon^{0123} = - \epsilon_{0123} = 1$, and repeated indices are summed over. Furthermore, we indicate operators by a hat, except for spinor and vector-field operators which are denoted by $\psi$ and $V^\mu$, respectively.
Throughout the paper, in order to distinguish quantities for Dirac and Proca fields, we will use the index $D$ or $P$, respectively.

\section{Interacting Dirac fields}
\label{intdirsec}

\subsection{Quantum transport for Dirac fields}

In this section we will briefly summarize the Wigner-function formalism derived in Refs. \cite{Weickgenannt:2020aaf,Weickgenannt:2021cuo} [see related work in Refs.~\cite{Fang:2016vpj,Florkowski:2018ahw,Gao:2019znl,Hattori:2019ahi,Wang:2019moi,Liu:2020flb,Manuel:2021oah}].
The Wigner function for spin-1/2 particles is defined as
\cite{DeGroot:1980dk,Heinz:1983nx,Vasak:1987um},
\begin{equation}
  \WD{\alpha\beta}(x,p)=\int \frac{d^4y}{(2\pi\hbar)^4} e^{-\frac i\hbar p\cdot y}
  \left\langle :\bar{\psi}_\beta\left(x_1\right)\psi_\alpha\left(x_2\right):\right\rangle \; ,
\end{equation}
with $x_{1,2}=x \pm y/2$ and $\psi (x)$ being the spinor field.
The Lagrangian density for Dirac fields is given by
\begin{equation}
\mathcal{L}_D=\bar{\psi}\left(\frac{i\hbar}{2}\gamma\cdot\olra{\partial}-m\right)\psi+\mathcal{L}_I \; , \label{LDint}
\end{equation}
with $\olra{\partial}\equiv \overrightarrow{\partial}-\overleftarrow{\partial}$ and  $\mathcal{L}_I$ being a general interaction Lagrangian, assumed to be a function only of spinors and their adjoints, but not of their derivatives, $\mathcal{L}_I=\mathcal{L}_I(\psi,\bar{\psi})$.
The equations of motion derived from the Lagrangian \eqref{LDint} read
\begin{subequations}
\label{dididirarac}
\begin{align}
 \left(i\hbar\gamma\cdot \partial-m\right)\psi(x)&=\hbar\rho(x)\; , \label{Dirac}\\
\bar{\psi}(x)\left(i\hbar\gamma\cdot \overleftarrow{\partial}+m\right)&=-\hbar\bar{\rho}(x)\; ,
\end{align}
\end{subequations}
where $\rho\equiv - (1/ \hbar) \partial \mathcal{L}_I /\partial \bar{\psi} $. From Eqs.~\eqref{dididirarac} one obtains the transport equation for the Wigner function \cite{DeGroot:1980dk},
\begin{equation}
 \left[ \gamma \cdot \left( p+\frac{i\hbar}{2} \partial \right) -m\right] W_D =\hbar\, \mathcal{C}\; , \label{Wignerkin}
\end{equation}
where
\begin{equation}
 \C_{\alpha\beta}\equiv \int \frac{d^4y}{(2\pi\hbar)^4} e^{-\frac i\hbar p\cdot y}
 \left\langle :\bar{\psi}_\beta(x_1)\rho_\alpha(x_2):\right\rangle \;.
\end{equation}
We decompose the Wigner function in terms of a basis of the generators of the Clifford algebra
\begin{equation}
W_D=\frac14\left(\F+i\gamma^5\Pc+\gamma \cdot \V+\gamma^5\gamma \cdot \A
+\frac12\sigma^{\mu\nu}\Sc_{\mu\nu}\right)\;, \label{dec}
\end{equation}
and substitute it into Eq.~\eqref{Wignerkin} to obtain the equations of motion for the coefficient functions~\cite{Weickgenannt:2020aaf}.
From the real part we find
 \begin{subequations}
 \label{real2}
\begin{eqnarray}
p\cdot \V -m\F&=& \hbar D_\F \; ,\label{F}\\
\frac{\hbar}{2}\partial\cdot  \A+m\Pc&=&-\hbar D_\Pc \;,\label{P}\\
p^\mu \F-\frac{\hbar}{2}\partial_\nu \Sc^{\nu\mu}-m\V^\mu&=& \hbar D^\mu_\V \; ,\label{V}\\
-\frac{\hbar}{2}\partial^\mu \Pc+\frac12\epsilon^{\mu\nu\alpha\beta}p_\nu S_{\alpha\beta}+m\A^\mu&=&-\hbar D^\mu_\A \; ,\label{A}\\
\frac{\hbar}{2}\partial^{[\mu} \V^{\nu]}-\epsilon^{\mu\nu\alpha\beta}p_\alpha \A_\beta-m\Sc^{\mu\nu}&=&\hbar D^{\mu\nu}_\Sc \; ,\label{SSS}
\end{eqnarray}
\end{subequations}
and from the imaginary part 
\begin{subequations}
\label{im2}
\begin{eqnarray}
\hbar\partial \cdot \V&=&2\hbar C_\F ,\label{Vkin}\\
p \cdot \A&=&\hbar C_\Pc \;,\label{orth}\\
\frac{\hbar}{2}\partial^\mu \F+p_\nu \Sc^{\nu\mu}&=&\hbar C_\V^\mu \;,\label{B}\\
p^{\mu}\Pc+\frac{\hbar}{4}\epsilon^{\mu\nu\alpha\beta}\partial_\nu \Sc_{\alpha\beta}&=& -\hbar C^\mu_\A \;,\label{Skin}\\
p^{[\mu} \V^{\nu]}+\frac{\hbar}{2}\epsilon^{\mu\nu\alpha\beta}\partial_\alpha \A_\beta&=&-\hbar C^{\mu\nu}_\Sc \;. \label{Akin}
\end{eqnarray}
\end{subequations}
Here we defined $D_i = \re \Tr\, (\tilde{\Gamma}_i \mathcal{C})$,
$C_i = \im \Tr\, (\tilde{\Gamma}_i \mathcal{C})$, $i = \F,\Pc,\V,\A,\Sc$, $\tilde{\Gamma}_\F=1$,
$\tilde{\Gamma}_\Pc =-i \gamma_5$, $\tilde{\Gamma}_\V = \gamma^\mu$,
$\tilde{\Gamma}_\A = \gamma^\mu \gamma^5$,
$\tilde{\Gamma}_\Sc= \sigma^{\mu \nu}$. The equations of motion \eqref{real2} and \eqref{im2} are solved employing an $\hbar$-gradient expansion \cite{Weickgenannt:2019dks,Weickgenannt:2021cuo}.

In quantum kinetic theory, it is convenient to introduce the phase-space spin variable $\ms^\mu$ and  define the distribution function as~\cite{Weickgenannt:2020aaf}
\begin{equation}
\f(x,p,\ms)\equiv \frac12\left[\F(x,p)-\hbar\delta V(x,p) -\ms\cdot \A(x,p)\right] \;, \label{fsdirac}
\end{equation}
where $\delta V$ is determined by
\begin{equation}
\label{Dmu}
D_\V^\mu=p^\mu \delta V+\mathcal{O}(\hbar) \;.
\end{equation}
Equation \eqref{Dmu}  holds if spin effects are considered to be of order $\mathcal{O}(\hbar)$, see Refs.~\cite{Weickgenannt:2020aaf,Weickgenannt:2021cuo} for details. Using the properties of the $\ms^\mu$-integration
\begin{eqnarray}
\label{fintprop}
\int \d S=2\;,\quad \int \d S \s^\mu \s^\nu =-2 P^{\mu\nu}\;,
\end{eqnarray}
with $P^{\mu\nu}\equiv g^{\mu\nu}-{p^\mu p^\nu}/{p^2}$ and $dS \equiv ({\sqrt{p^2}}/{\sqrt{3}\pi})  d^4\ms\,  \delta(\ms\cdot\ms+3)\delta(p\cdot \ms)$, one can prove that the functions $\F$, $\delta V$ and $\A^\mu$ are given by
\begin{eqnarray}
\int \d S\, \f=\F-\hbar\delta V\;,\quad \int \d S\, \s^\mu \f  =\A^\mu \;. \label{dsspin12}
\end{eqnarray}

The equation of motion for the distribution function has the form of a Boltzmann equation
\begin{equation}
p\cdot \partial \f= \mC[\f] \;, \label{boltz}
\end{equation}
where the collision term $\mC$ contains both local and nonlocal contributions~\cite{Weickgenannt:2020aaf,Weickgenannt:2021cuo}. In general, the distribution function $\f$ is not on-shell. However, it was shown in Refs.~\cite{Weickgenannt:2020aaf,Weickgenannt:2021cuo} that off-shell terms cancel on both sides of the Boltzmann equation \eqref{boltz}, and one is left with
\begin{equation}
\delta(p^2-m^2) p\cdot \partial f= \delta(p^2-m^2)\mC[f]\; , \label{boltzonshell}
\end{equation}
where $f$ is defined through
\begin{equation}
    \f=m\delta(p^2-M^2)f \;, \label{ff}
\end{equation}
with $M$ being an effective mass containing  interaction contributions. 

In order to solve the equations of motion \eqref{real2} and \eqref{im2}, we employ
an expansion in powers of $\hbar$ for the functions 
$\F, \mathcal{P}, \V^\mu, \A^\mu, \Sc^{\mu\nu}$ and
the collision terms $D_i, C_i$ [see, e.g., Refs.~\cite{Gao:2019znl,Weickgenannt:2019dks,Hattori:2019ahi,Weickgenannt:2020aaf}], e.g., for the scalar part 
\begin{equation}
	\F = \F^{(0)}+\hbar \F^{(1)} + \mathcal O (\hbar^2)\;.
	\end{equation}
Since gradients are always accompanied by factors of $\hbar$, 
this is effectively a gradient expansion.

\subsection{Canonical currents}
The so-called canonical energy-momentum and spin tensors are obtained from the interacting Dirac Lagrangian in \eq\eqref{LDint} using Noether's theorem \cite{Speranza:2020ilk}. 
 The canonical tensors are on the operator level given by
 \begin{subequations}
\begin{align}
\T_{D,C}^{\mu\nu}
&= \frac{i\hbar}{2}\psbar\gamma^\mu \olra{\partial}^\nu\ps - g^{\mu\nu}\mathcal{L}_D \; ,\label{tcan}\\
 \hbar\Sp_{D,C}^{\lambda,\mu\nu}
 &= \frac\hbar 4\psbar\{\gamma^\lambda,\sigma^{\mu\nu}\}\ps \n\\
 &=-\frac \hbar 2 \epsilon^{\lambda\mu\nu\alpha}\psbar\gamma_\alpha \gamma_5 \ps \;. \label{scan}
\end{align}
\end{subequations}
The normal-ordered ensemble averages
\begin{equation}
\TD{C}^{\mu\nu}\equiv \langle : \T_{D,C}^{\mu\nu}:\rangle\;, \quad
\SD{C}^{\lambda,\mu\nu}\equiv \langle: \Sp_{D,C}^{\lambda,\mu\nu}:\rangle
\end{equation}
 can be expressed in terms of the Wigner function as~\cite{Speranza:2020ilk}
 
\begin{subequations}
 \label{canonicalcurrents333}
\begin{eqnarray} 
\TD{C}^{\mu\nu}&=&\int d^4 p\, p^\nu \V^\mu \;,  \label{TmnDC}\\
 \SD{C}^{\lambda,\mu\nu}&=&-\frac{1}{2}\, \epsilon^{\lambda\mu\nu\alpha} \int d^4p\, \A_\alpha \;.  \label{scan33}
\end{eqnarray}
\end{subequations} 
Plugging \eq\eqref{SSS} into \eq\eqref{V} and then \eq\eqref{V} into \eq\eqref{TmnDC} we obtain, after considering spin effects to be of $\mathcal{O}(\hbar)$~\cite{Weickgenannt:2020aaf,Weickgenannt:2021cuo},
 \begin{subequations}
 \label{canonicalcurrents33}
\begin{eqnarray} 
\TD{C}^{\mu\nu}
&=& \int d\Gamma\,  p^\nu \left[ p^\mu +\frac{\hbar}{2}\Sigma_\ms^{\mu\lambda}\partial_\lambda+\frac{\hbar^2}{4m^2}\partial_\lambda(\partial^{\mu} p^\lambda-\partial^{\lambda}p^\mu)\right]
 f(x,p,\ms)+\frac{\hbar^2}{m}\int d^4p\, p^\nu D_\V^{(1)\mu}+\mathcal{O}(\hbar^3)\;,\label{tcan332}\\
 \SD{C}^{\lambda,\mu\nu} &=& 
 \frac{m^2}{2}  \int d\Gamma\, \frac{1}{p^2}\left(p^\lambda \Sigma_{\ms}^{\mu\nu}+p^\mu\Sigma_{\ms}^{\nu\lambda}+p^\nu\Sigma_{\ms}^{\lambda\mu}\right) f(x,p,\ms)\;, \label{scan332}
\end{eqnarray}
\end{subequations}
where we performed an expansion in $\hbar$ and defined $ d\Gamma \equiv d^4p\,  \delta(p^2-m^2)\, dS$ as well as the dipole-moment tensor
\begin{equation}
 \Sigma_{\ms}^{\mu\nu}\equiv -\frac1m \epsilon^{\mu\nu\alpha\beta}p_\alpha \ms_\beta\;.
\end{equation}
 Note that \eq\eqref{scan332} is exact at any order in the Planck constant \footnote{In \eqs\eqref{canonicalcurrents33} we do not take into account mass-shell corrections or the term proportional to $g^{\mu\nu}$ in the energy-momentum tensor. In general, such contributions can be nonvanishing in the presence of interactions, however, they can be neglected under the assumption of low density~\cite{DeGroot:1980dk}, which is employed in this work.}.
With the help of
 \eq\eqref{boltz}, we derive the following equations of motion,
 \begin{subequations}
\begin{align} \label{Tacan}
 \partial_\mu \TD{C}^{\mu\nu} &  = \int d\Gamma\, p^\nu\,  {\mC}[f] +\mathcal{O}(\hbar^2)=0 \;,\\
  \hbar\, \partial_\lambda \SD{C}^{\lambda,\mu\nu}
 & = \int d\Gamma\, \frac\hbar2 \left\{\Sigma_\ms^{\mu\nu}\, {\mC}[f]+p^{[\mu}\Sigma_\ms^{\nu]\lambda}\partial_\lambda f(x,p,\ms)\right\} = \TD{C}^{[\nu\mu]}\;. \label{total_conscan}
\end{align}
\end{subequations}
One can see from \eq\eqref{Tacan} that the fact that $p^\mu$ is a collisional invariant leads to the conservation of the energy-momentum tensor. Using \eq\eqref{Tacan} in \eq\eqref{tcan332}, we can express the canonical energy-momentum tensor as
\begin{align}
\label{TDCCD}
\TD{C}^{\mu\nu}=&\int d\Gamma\,   p^\nu \left[ p^\mu\left(1-\frac{\hbar^2}{4m^2}\partial^2\right)+\frac{\hbar}{2}\Sigma_\ms^{\mu\lambda}\partial_\lambda\right] f(x,p,\ms)+\frac{\hbar^2}{m}\int d^4p\, p^\nu D_\V^{(1)\mu}+\mathcal{O}(\hbar^3)\; .
\end{align}
Taking the antisymmetric part of \eq\eqref{TDCCD} and inserting it into \eq\eqref{total_conscan}, one can see that $\Sigma_\ms^{\mu\nu}$ is not conserved in a collision if and only if the interaction term $D_\V^\mu$ is nonzero. However, it can be seen from Eq.~\eqref{total_conscan} that the canonical spin tensor is not conserved even if $\Sigma_\ms^{\mu\nu}$ is a collisional invariant, and even if there are no interactions.
Furthermore, in the case of rigidly rotating global equilibrium, the canonical energy-momentum tensor is not symmetric either~\cite{Speranza:2020ilk}, cf.\ Section \ref{eomsec}. Therefore, the canonical spin tensor does not have a clear interpretation as a spin density, since the latter, in a physical picture, should change only through particle scatterings until the system is globally equilibrated.
At this point, we note that one can make use of the pseudo-gauge freedom in \eq\eqref{PGT} to obtain a set of energy-momentum and spin tensors with a clearer physical interpretation than the canonical ones.
 In the next sections, we will derive the so-called Hilgevoord-Wouthuysen, de-Groot-van-Leeuwen-van-Weert, and alternative Klein-Gordon currents, respectively, in the presence a general interaction term. 

\subsection{Hilgevoord-Wouthuysen currents}

A pseudo-gauge in which the energy-momentum tensor is symmetric for free fields, implying the conservation of the spin tensor, has been introduced by Hilgevoord and Wouthuysen (HW) in Refs.~\cite{HILGEVOORD19631,hilgevoord1965covariant}. The main idea of those works is to apply Noether's theorem to the Klein-Gordon Lagrangian for spinors, and then impose the Dirac equation as a subsidiary condition.
The pseudo-gauge potentials for the HW tensors in the free case read~\cite{Speranza:2020ilk}
\begin{subequations}
\label{potHW}
\begin{align}
  \hat{\Phi}\free^{\lambda,\mu\nu}&= \hat{M}^{[\mu\nu]\lambda}-g^{\lambda[\mu} \hat{M}_\rho^{\ \nu]\rho}\;, \label{pghw1}\\
   \hat{Z}\free^{\mu\nu\lambda\rho}&=-\frac{1}{8m}\psbar(\sigma^{\mu\nu}\sigma^{\lambda\rho}+\sigma^{\lambda\rho}\sigma^{\mu\nu})\ps \;, \label{pghw2}
\end{align}
\end{subequations}
where
\begin{equation}
 \hat{M}^{\lambda\mu\nu}\equiv \frac{i\hbar}{4m}\psbar\sigma^{\mu\nu}\olra{\partial}^\lambda \ps \;.
\end{equation}
For the interacting case, we consider the modifications of the potentials in \eqs\eqref{potHW} as follows
\begin{subequations}
\label{pghw34}
\begin{align}
  \hat{\Phi}_{HW}^{\lambda,\mu\nu}={}& \hat{M}^{[\mu\nu]\lambda}-g^{\lambda[\mu} \hat{M}_\rho^{\ \nu]\rho}+\hat{Q}^{\lambda\mu\nu} \;, \label{pghw3}\\
   \hat{Z}_{HW}^{\mu\nu\lambda\rho}={}&-\frac{1}{8m}\psbar(\sigma^{\mu\nu}\sigma^{\lambda\rho}+\sigma^{\lambda\rho}\sigma^{\mu\nu})\ps \;, \label{pghw4}
\end{align}
\end{subequations}
with
\begin{equation}
\hat{Q}^{\lambda\mu\nu}\equiv -\frac{\hbar}{4m}\bar{\rho}\gamma^\lambda\sigma^{\mu\nu}\psi-\frac{\hbar}{4m}\bar{\psi}\sigma^{\mu\nu}\gamma^\lambda\rho \;.
\end{equation}
In order to compute the interacting HW energy-momentum tensor $T_{D,HW}^{\mu\nu}$ from \eq\eqref{PGT}, we first consider the following part
\begin{align}
 &\TD{C}^{\mu\nu}-\hbar\partial_\lambda\left(M^{\nu\mu\lambda}+g^{\nu[\mu}M_\rho^{\ \lambda]\rho}\right)\n\\
 =& \int d^4p\, p^\nu \V^\mu-\frac{\hbar}{2m}\int d^4p\, \partial_\lambda \left(p^\nu \Sc^{\mu\lambda}+g^{\nu[\mu}\Sc^{\lambda]\rho}p_\rho\right)\n\\
=& \frac1m \int d^4p\, \left[p^\nu \left(p^\mu \F-\hbar D_\V^\mu\right)+\frac{\hbar^2}{4}\partial^\nu\left(\partial^\mu\F-2C_\V^\mu\right)  -\frac{\hbar^2}{4}g^{\mu\nu}\left( \partial^2 \F-2\, \partial\cdot C_\V \right) \right],
 \label{HWfromCwithM}
\end{align}
where \eqs\eqref{V} and \eqref{B} were inserted. The contribution due to the tensor $Q^{\lambda\mu\nu}$ to the energy-momentum tensor is given by 
\begin{align}
\partial_{\lambda}(Q^{\lambda\mu\nu}+Q^{\nu\mu\lambda}+Q^{\mu\nu\lambda})=& -\frac{\hbar}{4m}\partial_\lambda \left\langle: \left[\bar{\rho} \left(2ig^{\nu[\mu}\gamma^{\lambda]}+\epsilon^{\lambda\mu\nu\alpha}\gamma^5\gamma_\alpha\right) \psi+\bar{\psi} \left(-2ig^{\nu[\mu}\gamma^{\lambda]}+\epsilon^{\lambda\mu\nu\alpha}\gamma^5\gamma_\alpha\right) \rho\right]:\right\rangle\n\\
=& -\frac{\hbar}{m}\partial_\lambda g^{\nu[\mu}\im \left\langle:\bar{\psi} \gamma^{\lambda]}\rho:\right\rangle+\frac{\hbar}{2m}\epsilon^{\lambda\mu\nu\alpha}\partial_\lambda \re \left\langle:\bar{\psi}\gamma_\alpha \gamma^5 \rho:\right\rangle\n\\
=& -\frac{\hbar}{m}\int d^4p\,  \left(g^{\mu\nu}\partial\cdot C_\V-\partial^\nu C_\V^\mu-\frac12 \epsilon^{\lambda\mu\nu\alpha}\partial_\lambda D_{\A\alpha} \right)\; , \label{Qpartialcalc}
\end{align}
where we used the relation $\gamma^\lambda \sigma^{\mu\nu}=ig^{\lambda[\mu}\gamma^{\nu]}+\epsilon^{\lambda\mu\nu\rho}\gamma^5\gamma_\rho$.
Summing up \eqs\eqref{HWfromCwithM} and \eqref{Qpartialcalc} we find
\begin{align}
\label{THWWF}
\TD{HW}^{\mu\nu}=& \frac1m \int d^4p\, \left[p^\nu \left(p^\mu \F-\hbar D_\V^\mu\right)+\frac{\hbar^2}{4}(\partial^\nu\partial^\mu  - g^{\mu\nu} \partial^2)\F+\frac{\hbar^2}{4}\epsilon^{\lambda\mu\nu\alpha}\partial_\lambda D_{\A\alpha}\right] +\mathcal{O}(\hbar^3) \;.
\end{align}
We note that the antisymmetric part of the HW energy-momentum tensor arises solely from interactions. Considering \eq\eqref{Dmu}, one can see that this antisymmetric part is of second order in $\hbar$. This implies that the HW spin tensor is conserved in the absence of interactions.

We now give the explicit form of the HW spin tensor. 
Making use of the relation $\gamma^\lambda \gamma^\mu=g^{\lambda\mu}-i\sigma^{\lambda\mu}$, we can write the interacting Dirac equation and its adjoint \eqref{dididirarac} in the following form
\begin{subequations}
\label{dir23}
\begin{align}
i\hbar\partial^\lambda \ps =& -\hbar \sigma^{\lambda \mu}\partial_\mu \ps+m \gamma^\lambda\ps+\hbar \gamma^\lambda \rho \;, \\
-i \hbar\partial^\lambda \psbar=&-\hbar\partial_\mu\psbar \sigma^{\lambda\mu} +m \psbar \gamma^\lambda+\hbar\bar{\rho}\gamma^\lambda \;.
\end{align}
\end{subequations}
With the help of \eqs\eqref{dir23} we obtain a generalization of the Gordon decomposition~\cite{gordon1928strom} in the presence of a general interaction term, i.e.,
\begin{align}
\label{gordonn}
\psbar \gamma^\mu \ps = &\frac{i \hbar}{2m}\left[ \psbar \olra{\partial}^\mu \ps -i \left( \psbar \sigma^{\mu\nu}\partial_\nu \ps +\partial_\nu \psbar \sigma^{\mu\nu} \ps\right) \right] -\frac{\hbar}{2m}\left(\psbar \gamma^\lambda\rho+\bar{\rho}\gamma^\lambda \psi\right) \;.
\end{align}
The HW spin tensor is then found by applying a pseudo-gauge transformation with the potentials in \eq\eqref{pghw34} to the canonical spin tensor \eqref{scan} and using \eq\eqref{gordonn}:
\begin{widetext}
\begin{eqnarray}
 \hat{S}_{D,HW}^{\lambda,\mu\nu}&=& \frac14 \bar{\psi}\{\gamma^\lambda,\sigma^{\mu\nu}\}\psi+\frac{i\hbar}{4m}\left(\bar{\psi}\olra{\partial}^{[\nu}\sigma^{\mu]\lambda}\psi-g^{\lambda[\nu}\sigma^{\mu]\rho}\olra{\partial}_\rho \psi\right)-\frac{\hbar}{4m}\left[(\partial_\rho\bar{\psi}\sigma^{\lambda\rho}-\bar{\rho}\gamma^\lambda)\sigma^{\mu\nu}\psi+\bar{\psi}\sigma^{\mu\nu}(\sigma^{\lambda\rho}\partial_\rho \psi-\gamma^\lambda\rho)\right]\n\\
 && +\frac{\hbar}{8m}\bar{\psi}[\sigma^{\mu\nu},\sigma^{\lambda\rho}]\olra{\partial}_\rho \psi\n\\
 &=& \frac{i\hbar}{4m}\bar{\psi}\sigma^{\mu\nu}\olra{\partial}^\lambda \psi\;,
\end{eqnarray}
\end{widetext}
where also the identity
\begin{equation}
 [\sigma^{\mu\nu},\sigma^{\lambda\rho}]=2i(g^{\mu\rho}\sigma^{\nu\lambda}+g^{\nu\lambda}\sigma^{\mu\rho}-g^{\mu\lambda}\sigma^{\nu\rho}-g^{\nu\rho}\sigma^{\mu\lambda})
\end{equation}
was used. Performing the ensemble average and expressing the result in terms of the Wigner function we have
\begin{equation}
\label{Sflux}
S_{D,HW}^{\lambda,\mu\nu}=\frac{1}{2m}\int d^4p\, p^\lambda \Sc^{\mu\nu} \;.
\end{equation}
Putting everything together, we arrive at the HW tensors used in Ref.~\cite{Weickgenannt:2020aaf}, which read up to first order in $\hbar$
\begin{subequations}
\begin{eqnarray} \label{KleinGordontensors}
 \TD{HW}^{\mu\nu}&=&\!\!\!\int d\Gamma\,  p^\mu p^\nu f(x,p,\ms) + \mathcal{O}(\hbar^2)\;,\\
 \SD{HW}^{\lambda,\mu\nu}
  &=& \!\!\! \int d\Gamma\, p^\lambda\left(\frac{1}{2} \Sigma_\ms^{\mu\nu}-\frac{\hbar}{4m^2}
  p^{[\mu}\partial^{\nu]}\right) f(x,p,\ms) + \mathcal{O}(\hbar^2)\;, \label{SpinHW}
 \end{eqnarray}
\end{subequations}
where, in order to get \eq\eqref{SpinHW}, we made use of \eq\eqref{SSS}. As shown in Refs.~\cite{Weickgenannt:2020aaf,Weickgenannt:2021cuo}, the HW spin tensor is not conserved only in the presence of nonlocal particle scatterings.

\subsection{de Groot-van Leeuwen-van Weert and alternative Klein-Gordon currents}

\label{sec:glw_kg}

The energy-momentum and spin tensors used by de Groot, van Leeuwen and van Weert (GLW) in Ref.~\cite{DeGroot:1980dk} are equivalent to the HW currents up to first order in $\hbar$. They are derived from the canonical currents in the fully interacting case using a pseudo-gauge transformation with
\begin{subequations}
\label{pgtglw}
\begin{eqnarray}
\Phi^{\lambda,\mu\nu}_{GLW}
  &=& \frac{1}{2m}\int d^4p\, p^{[\mu} \Sc^{\nu]\lambda}\;, \\
Z_{GLW}^{\mu\nu\lambda\rho}&=&0 \;.
\label{pgtglw12}
\end{eqnarray}
\end{subequations}
Following similar steps as in the HW case, we obtain from \eq\eqref{PGT}
\begin{subequations}
\begin{align}
\TD{GLW}^{\mu\nu}={}& \int d^4p\, p^\nu\left( \V^\mu+\frac{\hbar}{2}  \partial_\lambda\Sc^{\mu\lambda}\right)\n\\
={}&\frac 1m \int d^4p\, p^\nu \left(p^\mu  \F-\hbar D_\V^\mu\right)\;, \label{tglwwfff}\\
\SD{GLW}^{\lambda,\mu\nu}={}&-\frac{1}{2}\, \epsilon^{\lambda\mu\nu\alpha} \int d^4p\, \A_\alpha-\frac{1}{2m}\int d^4p\, p^{[\nu} \Sc^{\mu]\lambda}\n\\
={}&\frac{1}{2m}\int d^4p\, \left[p^\lambda \Sc^{\mu\nu}-\hbar\epsilon^{\lambda\mu\nu\alpha}\left(\frac12\partial_\alpha\Pc-D_{\A\alpha}\right)\right] \;, \label{sglwwfff}
\end{align}
\end{subequations}
where in the last equality we used Eq.~\eqref{A}. We see that, since $\Pc$ and $D_\A^\alpha$ have contributions starting at first order in $\hbar$~\cite{Weickgenannt:2020aaf}, the HW and GLW currents differ only at second and higher orders in $\hbar$. Note that, unlike in the HW spin tensor \eqref{Sflux}, the GLW spin tensor is not expressed only by the flux of $\Sc^{\mu\nu}$. Furthermore, the term with $\partial_\alpha\Pc$ is separately conserved and hence does not enter the equation of motion for the spin tensor.
Modifying the GLW pseudo-gauge transformations \eqref{pgtglw} by only adding
\begin{equation}
Z_{KG}^{\mu\nu\lambda\rho} =  \frac{1}{4m}\epsilon^{\mu\nu\lambda\rho}\int d^4p\, \Pc
\end{equation}
to \eq\eqref{pgtglw12}, we can remove the term containing $\partial_\alpha \Pc$ from the GLW spin tensor without affecting the GLW energy-momentum tensor \eqref{tglwwfff} 
[alternatively, we could also add $-{\hbar}/({2m})\epsilon^{\lambda\mu\nu\alpha}\partial_\alpha \Pc$ to $\Phi^{\lambda,\mu\nu}$]. In this case, we obtain the currents   corresponding to the alternative Klein-Gordon (KG) pseudo-gauge~\cite{Speranza:2020ilk} with the spin tensor given by
\begin{equation}
 \Sp_{D,KG}^{\lambda,\mu\nu}
=\frac{i\hbar}{4m}\bar{\psi}\sigma^{\mu\nu}\olra{\partial}^\lambda\psi+\frac{\hbar}{2m}\epsilon^{\lambda\mu\nu\rho} \re \bar{\psi}\gamma_\rho\gamma^5\rho\; ,
\end{equation}
which can be expressed in terms of the components of the Wigner function as
\begin{equation}
\SD{KG}^{\lambda,\mu\nu}=\frac{1}{2m}\int d^4p\, \left(p^\lambda\Sc^{\mu\nu}+\hbar\epsilon^{\lambda\mu\nu\rho} D_{\A\rho}\right)\; .
 \label{snewint}
\end{equation}

\section{Free Proca fields}
\label{freeprocsec}

In contrast to the case of spin-1/2 particles, there has been only little work on the spin tensor for Proca fields up
to now. For this reason, we start with a general discussion of different pseudo-gauges for free, massive spin-1 fields,
pointing out the analogies to Dirac fields.

\subsection{Canonical currents}

We consider the Lagrangian of a free complex Proca field $V^\mu$ given as
\begin{equation}
  \mathcal{L}_{P0}=\hbar \left(-\frac{1}{2}V^{\dagger\mu\nu} V_{\mu\nu} +\frac{m^2}{\hbar^2}V^{\dagger\mu} V_\mu \right)\;,
\label{action}
\end{equation}
where $V^{\mu\nu}\equiv\partial^{[\mu} V^{\nu]}$ is the field-strength tensor.
This Lagrangian generates the following equations of motion for the Proca fields
\begin{equation}
    \hbar^2\partial_\mu V^{\mu\nu}+m^2 V^\nu=0\; , \label{procaeq}
\end{equation}
from which the constraint equation 
\begin{equation}
\label{constr_V}
\partial \cdot V = 0
\end{equation}
follows by taking the divergence.

The invariance of the action associated to the Lagrangian \eqref{action} under spacetime translations and Lorentz transformations implies the conservation of the canonical energy-momentum and total angular momentum tensors $\hat{T}_{P,C}^{\mu\nu}$ and $\J_{P,C}^{\lambda,\mu\nu}$, respectively.
These quantities read
\begin{subequations} \label{canproca}
\begin{eqnarray}
\T_{P,C}^{\mu\nu}&=&{-}\hbar \left(V^{\dagger\mu\rho}\partial^\nu V_\rho+V^{\mu\rho}\partial^\nu V^{\dagger}_\rho    \right)-g^{\mu\nu} \mathcal{L}_{P0}\;,\label{T_C}\\
\J_{P,C}^{\lambda,\mu\nu}&=& x^{\mu}\T_{P,C}^{\lambda\nu} - x^{\nu}\T_{P,C}^{\lambda\mu} +\hbar\, \Sp_{P,C}^{\lambda,\mu\nu}\;,\label{J_C}
\end{eqnarray}
\end{subequations}
with 
\begin{equation}
\S_{P,C}^{\lambda,\mu\nu}\equiv  V^{\dagger\lambda[\nu}V^{\mu]}+V^{\lambda[\nu}V^{\dagger\mu]}\; .
\end{equation}
As for the spin-1/2 case, the canonical spin tensor for free spin-1 particles is not conserved, as the energy-momentum tensor \eqref{T_C} is not symmetric, leading to the same problems as discussed above.

Following Refs.~\cite{Vasak:1987um, Elze:1986hq,Elze:1989un,Huang:2020kik,Hattori:2020gqh},
we define the massive spin-1 Wigner function as
\begin{equation}
\label{Wigner_function}
W_P^{\mu\nu}(x,p)\equiv-\frac{2}{\hbar(2\pi\hbar)^4} \int \d^4 v\, e^{-ip\cdot v/\hbar} \avg{V^{\dagger\mu}\left(x+\frac v2\right) V^\nu\left(x-\frac v2\right)}\;.
\end{equation}
In terms of the Wigner function \eqref{Wigner_function}
 we can express \eqs\eqref{canproca} as
\begin{eqnarray}
\TP{C}^{\mu\nu}&=&\int \d^4 p \left[\left(p^\mu p^\nu +\frac{\hbar^2}{4}\partial^\mu \partial^\nu\right)\mathrm{Tr}\,W_P-\left(p^\nu p_\rho +\frac{\hbar^2}{4}\partial^\nu \partial_\rho \right)\WP{S}^{\rho\mu}-\frac{i\hbar}{2} p^{[\nu} \partial_{\rho]} \WP{A}^{\rho\mu}\right]-g^{\mu\nu} \left\langle: \mathcal{L}_{P0}:\right\rangle \;,\\
\SP{C}^{\lambda,\mu\nu}&=&i\int \d^4 p\, \Big(2 p^\lambda \WP{A}^{\mu\nu}+p^{[\mu} \WP{A}^{\nu]\lambda} -\frac{i\hbar}{2} \partial^{[\nu} \WP{S}^{\mu]\lambda}  \Big)\;,\label{S_C_Wigner}
\end{eqnarray}
where we defined the symmetric part $\WP{S}^{\mu\nu}\equiv (1/2) W_P^{(\mu\nu)}$ as well as the antisymmetric part $\WP{A}^{\mu\nu}\equiv (1/2) W_P^{[\mu\nu]}$ of the Wigner function. 

Using the Proca equation \eqref{procaeq} and the constraints on the Wigner function
\begin{equation}
p_\mu \WP{S}^{\mu\nu}-\frac{i\hbar}{2}\partial_\mu \WP{A}^{\mu\nu} =
p_\mu \WP{A}^{\mu\nu}-\frac{i\hbar}{2}\partial_\mu \WP{S}^{\mu\nu}=0\;,
\end{equation}
which follow from \eq\eqref{constr_V}, one can rewrite the canonical energy-momentum as
\begin{eqnarray}
\TP{C}^{\mu\nu}&=&\int \d^4 p \left[\left(p^\mu p^\nu +\frac{\hbar^2}{4}\partial^\mu \partial^\nu\right)\mathrm{Tr}\,W_P-\hbar^2\frac12\partial^\nu \partial_\rho \WP{S}^{\rho\mu}-{i\hbar} p^{\nu} \partial_{\rho} \WP{A}^{\rho\mu}-g^{\mu\nu}\frac{\hbar^2}{4}\left( \partial^2 \mathrm{Tr}\, W_P-\partial_\lambda\partial_\rho W^{\lambda\rho}_P\right) \right]. \;\;\;\;
\end{eqnarray}
As expected, $\TP{C}^{\mu\nu}$ approaches the classical symmetric form in the limit $\hbar\rightarrow0$.

The definition of the energy-momentum and spin tensors can be changed by applying the pseudo-gauge transformations \eqref{PGT}.
For instance, applying a Belinfante pseudo-gauge transformation~\cite{belinfante1939spin} with $\Phi_B^{\lambda,\mu\nu}=\SP{C}^{\lambda,\mu\nu}, Z^{\mu\nu\lambda\rho}_B=0$ yields
\begin{subequations}
\begin{eqnarray}
\TP{B}^{\mu\nu}&=&\hbar \left\langle:\left[V^{\mu\rho}V_\rho^{\dagger\;\nu}+V^{\dagger\mu\rho}V_\rho^{\;\;\nu} +\frac{m^2c^2}{\hbar^2}\left(V^{\dagger\mu}V^\nu+V^\mu V^{\dagger\nu}\right)\right]:\right\rangle-g^{\mu\nu}\avg{\mathcal{L}_{P0}}\n\\
&=&   \int \d^4 p \left[\left(p^\mu p^\nu +\frac{\hbar^2}{4}\partial^\mu \partial^\nu\right)\mathrm{Tr}\,W_P-\hbar^2\frac12\partial^{(\nu} \partial_\rho \WP{S}^{\mu)\rho}+{i\hbar} p^{(\nu} \partial_{\rho} \WP{A}^{\mu)\rho}+\frac12\hbar^2 \partial^2 \WP{S}^{\mu\nu}\right.\n\\
&&\left.-g^{\mu\nu}\frac{\hbar^2}{4}\left( \partial^2 \mathrm{Tr} \,W_P-\partial_\lambda\partial_\rho W^{\lambda\rho}_P\right) \right]\;, \\
\SP{B}^{\lambda,\mu\nu}&=&0 \;,
\end{eqnarray}
\end{subequations}
where we also made use of the equations of motion.

\subsection{Hilgevoord-Wouthuysen currents}

Following the idea by Hilgevoord and Wouthuysen~\cite{HILGEVOORD19631}, we find a set of symmetric energy-momentum tensor and conserved spin tensor for free fields by deriving the conserved currents from the Lagrangian
\begin{equation}
\label{FermiL}
\mathcal{L}'_{P}\equiv-\hbar \left[ \left(\partial_\mu V_{\nu}^\dagger\right)\partial^\mu V^\nu-(\partial\cdot V^\dagger)\partial\cdot V -\frac{m^2}{\hbar^2} V^{\dagger\mu}V_\mu\right]\;,
\end{equation}
which differs from $\mathcal{L}_{P0}$ by a total divergence and thus yields the same equations of motion. Applying Noether's theorem to the Lagrangian \eqref{FermiL} we obtain
\begin{subequations}
\begin{eqnarray}
\T_{P,HW}^{\mu\nu}&=&-\hbar  \left[\left(\partial^\mu V^\lambda\right)\partial^\nu V^\dagger_\lambda  +\left(\partial^\mu V^{\dagger\lambda}\right)\partial^\nu V_\lambda\right] -g^{\mu\nu} \mathcal{L}'_\mathrm{P} \label{T_HW}\;,\\
\Sp_{P,HW}^{\lambda,\mu\nu}&=&-  \left[\left(\partial^\lambda V^{\dagger[\mu}\right)V^{\nu]}+\left(\partial^\lambda V^{[\mu}\right)V^{\dagger\nu]}\right]\label{S_HW}\;.
\end{eqnarray}
\end{subequations}
The spin tensor $\Sp_{P,HW}^{\lambda,\mu\nu}$ is conserved since the energy-momentum tensor is symmetric, implying that the global spin $\Sp_{P,HW}^{\mu\nu}$ transforms as a tensor \cite{Speranza:2020ilk}.

In analogy to the spin-1/2 case, we can relate the HW currents to the pseudo-gauge transformation
\begin{subequations}\label{HW_free_Proca}
\begin{eqnarray}
\hat{\Phi}^{\lambda\mu\nu}_{HW,\mathrm{free}}&=&  \hat{M}^{[\mu\nu]\lambda}-g^{\lambda[\mu} \hat{M}_\rho^{\;\;\nu]\rho}  \;,\\
\hat{Z}^{\mu\nu,\lambda\rho}_{HW,\mathrm{free}}&=&-\frac{1 }{2} \left(V^{\dagger[\mu} g^{\nu][\lambda} V^{\rho]}+\mathrm{h.c.}\right)\;,
\end{eqnarray}
\end{subequations}
where h.c.\ stands for the hermitian conjugate and
\begin{equation}
 \hat{M}^{\lambda\mu\nu}\equiv\frac{1}{2} \left(V^{\dagger\mu} \odouble{\partial}^\lambda V^\nu+\mathrm{h.c.}\right)=\frac{1}{2}\left( V^{\dagger\mu} \partial^\lambda V^\nu -V^{\dagger\nu} \partial^\lambda V^\mu+\mathrm{h.c.} \right)\; .
\end{equation}
When performing the pseudo-gauge transformation, one also makes use of the equations of motion.

The HW currents in terms of the Wigner function are given by
\begin{subequations}
\begin{eqnarray}
\TP{HW}^{\mu\nu}
&=& \int \d^4 p \left[p^\mu p^\nu +\frac{\hbar^2}{4}\left(\partial^\mu \partial^\nu-g^{\mu\nu}\partial^2\right)\right]\mathrm{Tr}\,W_P \;,\\
\SP{HW}^{\lambda,\mu\nu}
&=&i\int \d^4 p\, p^\lambda W^{[\mu\nu]}_P\label{S_HW_Wigner}\;.
\end{eqnarray}
\end{subequations}
Identifying $\Tr\, W_P$ with the scalar distribution $\F$ and $W^{[\mu\nu]}_P$ with the dipole moment $\Sc^{\mu\nu}$, these expressions are formally equivalent to the HW currents in terms of the Wigner function for spin 1/2 in the free case~\cite{Speranza:2020ilk} [cf. \eqs\eqref{THWWF} and \eqref{Sflux}].

\subsection{Alternative Klein-Gordon currents}

One can also obtain a set of symmetric energy-momentum tensor and conserved spin tensor considering the alternative Klein-Gordon Lagrangian analogously to the case of spin-1/2 particles, 
\begin{equation}
\mathcal{L}_{P,KG}^\prime= -\hbar \left[-\frac12 \left(V^\mu \partial^2 V^\dagger_\mu+V^\dagger_\mu \partial^2 V^\mu\right)-(\partial\cdot V^\dagger)\partial\cdot V-\frac{m^2}{\hbar^2} V^{\dagger\mu}V_\mu  \right] \; ,
\end{equation}
which differs from \eq\eqref{FermiL} by a total divergence and hence also yields the same equations of motion. The resulting set of tensors reads
\begin{subequations}\label{T_KG}
\begin{align}
\T_{P,KG}^{\mu\nu}
=&\frac{\hbar }{2}V^\lambda \olra{\partial}^\mu \olra{\partial}^\nu V^\dagger_\lambda \;, \\
\Sp_{P,KG}^{\lambda,\mu\nu}={}& \Sp_{P,HW}^{\lambda,\mu\nu} \; ,
\end{align}
\end{subequations}
where we used $\mathcal{L}^\prime_{P,KG}=0$ after imposing the equations of motion.
One can obtain these currents from the canonical ones by employing a pseudo-gauge transformation with
\begin{subequations}\label{pgtkgproca}
\begin{align}
\hat{\Phi}^{\lambda,\mu\nu}_{KG,\mathrm{free}}={}&  g^{\lambda[\nu} V^{\mu]\rho} V^\dagger_\rho- V^{\dagger\lambda} V^{\mu\nu}- V^{\dagger[\mu} \partial^{\nu]} V^\lambda+ \text{h.c.}\; , \\
\hat{Z}^{\mu\nu\lambda\rho}_{KG,\mathrm{free}}={}& -\frac{1}{2} \left(V^{\dagger[\mu} g^{\nu][\lambda} V^{\rho]}+\mathrm{h.c.}+\frac12 V^{\dagger\beta} V_\beta g^{[\nu}_\alpha g^{\mu][\lambda}g^{\rho]\alpha}  \right)
\end{align}
\end{subequations}
and using the equations of motion.
We can express the KG energy-momentum tensor in terms of the Wigner function as 
\begin{equation}
\TP{KG}^{\mu\nu}=\int \d^4 p\, p^\mu p^\nu \mathrm{Tr}\,W_P\label{T_KG_Wigner}\;.
\end{equation}
Thus, we have found a pair of spin and energy-momentum tensors that can be represented as moments of the scalar distribution function $\mathrm{Tr}\,W_P$ and the antisymmetric part $W^{[\mu\nu]}_P$, closely mimicking the Klein-Gordon currents in the spin-1/2 theory, see Section \ref{sec:glw_kg} and Ref.~\cite{Speranza:2020ilk}.

\section{Interacting Proca fields}
\label{intprocsec}

\subsection{Quantum transport for Proca fields}
\label{sec:pr_int}

In the interacting case, we consider a Lagrangian which is given as the sum of the free Proca Lagrangian \eqref{action} and  a general interaction term $\mathcal{L}_\text{int}$, which we assume to be independent of the derivatives of the Proca field,
\begin{equation}
\label{Lagrangian}
\mathcal{L}_P=-\hbar \left(\frac{1}{2}V^{\dagger\mu\nu} V_{\mu\nu} -\frac{m^2}{\hbar^2}V^{\dagger\mu}V_\mu\right)+\mathcal{L}_\text{int}\;.
\end{equation}
The equations of motion now read
\begin{equation}
\left(\partial^2 +\frac{m^2}{\hbar^2}    \right)V^\mu-\partial^\mu \partial\cdot V =\rho^\mu\;,
\label{EoM}
\end{equation}
where we defined $\rho^\mu \equiv-({1}/{\hbar }) {\partial \mathcal{L}_\text{int}}/{\partial V^\dagger_\mu}$. Taking the divergence of \eq\eqref{EoM} we obtain the new constraint equation
\begin{equation}
\label{constraint}
\partial \cdot V= \frac{\hbar^2}{m^2} \partial\cdot \rho \;.
\end{equation}
In this section, we consider a general interaction which does not involve gauge fields so that we can stick to the definition of the Wigner function in  \eq\eqref{EoM}. In the case where the massive vector particles interact with an electromagnetic field the Wigner function has to be defined in a gauge-invariant way, see Section \ref{emfsec}. The equations of motion take the form
\begin{equation}
\left( p^2-m^2-\frac{\hbar^2}{4}\partial^2+i\hbar p\cdot \partial   \right)W_P^{\mu\nu} - \frac{\hbar}{m^2}\left(p^\nu p_\alpha -\frac{\hbar^2}{4}\partial^\nu \partial_\alpha +\frac{i\hbar}{2}p^{(\nu}\partial_{\alpha)}\right) C^{\mu\alpha}=-\hbar C^{\mu\nu}\;,
\label{eom567}
\end{equation}
while from \eq\eqref{constraint} we derive the constraint equations
\begin{subequations} \label{boppconsint}
\begin{eqnarray}
\left(p_\mu +\frac{i\hbar}{2}\partial_\mu\right) W_P^{\nu\mu}&=&\frac{\hbar}{m^2} \left(p_\mu +\frac{i\hbar}{2}\partial_\mu\right) C^{\nu\mu}\;,\\
\left(p_\mu -\frac{i\hbar}{2}\partial_\mu\right) W_P^{\mu\nu}&=&\frac{\hbar}{m^2}\left(p_\mu -\frac{i\hbar}{2}\partial_\mu\right) C^{*\nu\mu} \;.
\end{eqnarray}
\end{subequations}
Here we employed the relations
\begin{subequations}
\begin{eqnarray}
\label{D}
\left(p^\mu +\frac{i\hbar}{2}\partial^\mu\right) W_P^{\alpha\beta}(x,p)&=&-2i \frac{1}{(2\pi\hbar)^4} \int \d^4 v e^{-ip\cdot v/\hbar} \avg{V^{\dagger\alpha}(x_1) \partial^\mu V^\beta(x_2)}\;,\\
\left(p^\mu -\frac{i\hbar}{2}\partial^\mu\right) W_P^{\alpha\beta}(x,p)&=&2i \frac{1}{(2\pi\hbar)^4} \int \d^4 v e^{-ip\cdot v/\hbar} \avg{[\partial^\mu V^{\dagger\alpha}(x_1)] V^\beta(x_2)}\;,
\label{Dstar}
\end{eqnarray}
\end{subequations}
 used the fact that the Wigner function is hermitian, and defined
\begin{equation}
\label{eq:coll_int_1}
C^{\mu\nu}\equiv-\frac{2}{(2\pi\hbar)^4} \int \d^4 y\, e^{-ip\cdot y/\hbar} \avg{V^{\dagger\mu}(x_1)\rho^\nu(x_2) }\; .
\end{equation}
Similarly, we define the hermitian objects
\begin{equation}
\delta M^{\mu\nu}\equiv -\frac{1}{2}\left( C^{\mu\nu}  +C^{*\nu\mu}   \right)\; ,\qquad
\C^{\mu\nu}\equiv \frac{i}{2}\left( C^{\mu\nu} -C^{*\nu\mu}   \right)\; .
\end{equation}
Splitting both the Wigner function $W_P^{\mu\nu}$ and the collision terms $\delta M^{\mu\nu},\C^{\mu\nu}$ into symmetric and antisymmetric parts, we can add and subtract the constraint equations \eqref{boppconsint} to obtain
\begin{subequations}\label{constr2}
\begin{eqnarray}
p_\mu \WP{S}^{\mu\nu}-\frac{i\hbar}{2}\partial_\mu \WP{A}^{\mu\nu} &=& \frac{\hbar}{m^2}\left[p_\mu \left(i\C^{\mu\nu}_A-\delta M^{\mu\nu}_S\right)+\frac{\hbar}{2}\partial_\mu\left(\C^{\mu\nu}_S+i\delta M^{\mu\nu}_A\right)   \right]\;,\\
p_\mu \WP{A}^{\mu\nu}-\frac{i\hbar}{2}\partial_\mu \WP{S}^{\mu\nu}&=&\frac{\hbar}{m^2}\left[p_\mu \left(i\C^{\mu\nu}_S-\delta M^{\mu\nu}_A\right)+\frac{\hbar}{2}\partial_\mu\left(\C^{\mu\nu}_A+i\delta M^{\mu\nu}_S\right)   \right]\;.
\end{eqnarray}
\end{subequations}
It should be noted that the symmetric parts of $\delta M^{\mu\nu}$ and $\C^{\mu\nu}$ are real, while their antisymmetric parts are imaginary.
Furthermore, from \eq\eqref{eom567} we derive the Boltzmann-like equation for the Wigner function
\begin{eqnarray}
p\cdot \partial W_P^{\mu\nu}&=&\C^{\mu\nu}-\frac{1}{2m^2}\left[ \left(p^\nu p_\alpha -\frac{\hbar^2}{4}\partial^\nu \partial_\alpha +\frac{i\hbar}{2}p^{(\nu}\partial_{\alpha)}\right) \left(\C^{\mu\alpha} -i\delta M^{\mu\alpha}  \right)+\text{h.c.}  \right]\label{Boltzmann} \;.
\end{eqnarray}
Splitting into symmetric and antisymmetric part, we find
\begin{subequations}
\begin{eqnarray}
p\cdot \partial \WP{S}^{\mu\nu}&=&\C^{\mu\nu}_S-\frac{1}{2m^2}\left[ \left(p_\alpha p^{(\mu}-\frac{\hbar^2}{4}\partial_\alpha \partial^{(\mu}\right)\left(\C^{\nu)\alpha}_S-i\delta M_A^{\nu)\alpha}\right) +\frac{\hbar}{2}\left(p_\alpha \partial^{(\mu}+\partial_\alpha p^{(\mu}\right)\left(i\C_A^{\nu)\alpha}+\delta M^{\nu)\alpha}_S\right)  \right]\label{Boltzmann_S} \;,\qquad\\
p\cdot \partial \WP{A}^{\mu\nu}&=&\C^{\mu\nu}_A-\frac{1}{2m^2}\left[\left(p_\alpha p^{[\mu}-\frac{\hbar^2}{4}\partial_\alpha \partial^{[\mu}\right)\left(i\delta M_S^{\nu]\alpha}-\C^{\nu]\alpha}_A\right) -\frac{\hbar}{2}\left(p_\alpha \partial^{[\mu}+\partial_\alpha p^{[\mu}\right)\left(i\C_S^{\nu]\alpha}+\delta M^{\nu]\alpha}_A\right)  \right]\label{Boltzmann_A}\;.
\end{eqnarray}
\end{subequations}

In the following we decompose the Wigner function and all related quantities with respect to the four-momentum $p^\mu$,
\begin{subequations}\label{decomp_1}
\begin{eqnarray}
\WP{S}^{\mu\nu}&=&E^{\mu\nu} f_E + \frac{p^{(\mu}}{2p}F_S^{\nu)}+F_P^{\mu\nu}+P^{\mu\nu}f_P\;,\\
\WP{A}^{\mu\nu}&=&i\frac{p^{[\mu}}{2p}F_A^{\nu]}+i\epsilon^{\mu\nu\alpha\beta} \frac{p_\alpha}{m}G_\beta\;,\\
\C^{\mu\nu}_S&=&E^{\mu\nu} \C_E + \frac{p^{(\mu}}{2}\C_S^{\nu)}+\C_P^{\mu\nu}+P^{\mu\nu}\C_P\;,\\
\C^{\mu\nu}_A&=&i\frac{p^{[\mu}}{2p}\C_A^{\nu]}+i\epsilon^{\mu\nu\alpha\beta} \frac{p_\alpha}{m}\C_{G,\beta}\;,\\
\delta M^{\mu\nu}_S&=&E^{\mu\nu} \D_E + \frac{p^{(\mu}}{2p}\D_S^{\nu)}+\D_P^{\mu\nu}+P^{\mu\nu}\D_P\;,\\
\delta M^{\mu\nu}_A&=&i\frac{p^{[\mu}}{2p}\D_A^{\nu]}+i\epsilon^{\mu\nu\alpha\beta} \frac{p_\alpha}{m}\D_{G,\beta}\;,
\end{eqnarray}
\end{subequations}
with $p\equiv\sqrt{p^2}$, $E^{\mu\nu}\equiv {p^\mu p^\nu}/{p^2}$, and $F_S\cdot p=F_A\cdot p=G\cdot p=0$, $F_P^{\mu\nu}p_\nu=0$, with $F_P^{\mu\nu}$ symmetric and traceless. Analogous proporties hold for the components of the collision terms $\C^{\mu\nu}$, $\delta M^{\mu\nu}$.
The constraint equations \eqref{constr2} then determine the Wigner-function components $f_E$, $F_S^\mu$, and $F_A^\mu$ in terms of $f_P$, $F_P^{\mu\nu}$, and $G^\mu$. Using the definition of the collision term \eqref{eq:coll_int_1} and the constraint \eqref{constraint}, we obtain
\begin{equation}
   \left(p_\mu -\frac{i\hbar}{2}\partial_\mu\right) C^{\mu\nu}=\mathcal{O}(\hbar)\; ,\qquad \left(p_\mu +\frac{i\hbar}{2}\partial_\mu\right) C^{\ast\mu\nu}=\mathcal{O}(\hbar)\; ,
\end{equation}
from which it follows that $\C_E^{(0)}=\D_E^{(0)}=0$.
As done in the spin-1/2 case, we consider a situation in which polarization effects arise only at first or higher order in $\hbar$. This implies that we do not have any vector or tensor anisotropy at zeroth order, i.e., $G^{(0)}_\mu=0$ and $F_P^{(0)\mu\nu}=0$. Following the same logic as explained in Ref.~\cite{Weickgenannt:2020aaf} and considering the quantum numbers, vectors and tensors at our disposal, we conclude $\C^{(0)\mu}_S=\C^{(0)\mu}_A=\D^{(0)\mu}_S=\D^{(0)\mu}_A=0$. Under this assumption, we obtain from the real parts of \eqs\eqref{constr2}
\begin{subequations}\label{constr_coll}
\begin{eqnarray}
f_E&=&\frac{\hbar^2}{4p^2} P^{\alpha\beta} \partial_\alpha \partial_\beta f_P^{(0)}-\frac{\hbar}{m^2}\D_E+\mathcal{O}(\hbar^3)\;,\\
F_S^{\nu}&=&\mathcal{O}(\hbar^2)\;,\\
p F_A^{\nu}&=&\hbar P^{\nu\mu}\partial_\mu f_P^{(0)}+\mathcal{O}(\hbar^2)\;.
\label{81c}
\end{eqnarray}
\end{subequations}
Furthermore, we derive from \eq\eqref{Boltzmann} the following Boltzmann-like equations of motion for the independent components,
\begin{subequations}
\label{compbolbol}
\begin{eqnarray}
p\cdot \partial f_P&=&\C_P+\mathcal{O}(\hbar^2)\; , \label{Boltzmann_scalar_1}\\
p\cdot \partial F_P^{\mu\nu}&=& \C^{\mu\nu}_P+\mathcal{O}(\hbar^2)\; ,\label{Boltzmann_tensor_1}\\
p\cdot \partial G^{\mu}&=& \C_G^{\mu}+\mathcal{O}(\hbar^2)\label{Boltzmann_vector_1}\;.
\end{eqnarray}
\end{subequations}

Analogously to the distribution function \eqref{fsdirac} in extended phase space for spin-1/2 particles, we define the spin-1 distribution function as
\begin{equation}
\f(x,p,\s)\equiv f_P-\s\cdot G+\frac54\,\s^{\mu}\s^{\nu} F_{P,\mu\nu}   \;. \label{fsproca}
\end{equation}
We note that for massive spin-1 particles, the number of degrees of freedom determining the spin state is larger than that for spin-1/2 particles. In fact, in addition to the usual vector polarization, we also have  spin degrees of freedom which are called tensor polarization~\cite{Leader:2001}.
The last term in \eq\eqref{fsproca}, which is absent for Dirac particles, precisely describes the additional degrees of freedom due to tensor polarization \cite{Leader:2001}.
In the spin-1 case, it is convenient to define the measure in spin space as $dS\equiv (3\sqrt{p^2}/2\sqrt{2}\pi)d^4 \s \delta(\s\cdot \s+2)\delta(p\cdot \s)$, such that
\begin{equation}
\label{spin1fintprop}
\int d S =3 \;,\quad \int dS \s^\mu \s^\nu =-2P^{\mu\nu}\;,\quad
\int \d S P^{\mu\nu}_{\rho\sigma}\s^{\rho} \s^{\sigma} \s_\alpha \s_\beta =\frac{8}{5}P^{\mu\nu}_{\alpha\beta}\;,
\end{equation}
where we defined $P^{\mu\nu}_{\alpha\beta}\equiv[(1/2) P^{(\mu}_\alpha P^{\nu)}_\beta-(1/3) P^{\mu\nu} P_{\alpha\beta}]$, cf. the spin-1/2 case in \eq\eqref{fintprop}. Using Eq. \eqref{spin1fintprop}, we obtain the independent components of the Wigner function from the distribution function as
\begin{eqnarray}
\int \d S\, \f=3f_P\;,\quad \int \d S\, \s^\mu \f  =2\,G^\mu \;,\quad \int \d S\,P^{\mu\nu}_{\alpha\beta}\s^{\alpha} \s^{\beta} \f =2F_P^{\mu\nu}\;. \label{fkgfkf}
\end{eqnarray}
Making use of \eqs\eqref{compbolbol}, we find the Boltzmann equation for the spin-1 Wigner function to be
\begin{equation}
p\cdot \partial \f =\mathfrak{C}[\f]\;,
\end{equation}
where 
\begin{eqnarray}
\mathfrak{C}[\f]&\equiv& \C_P -\,\s\cdot \C_G +\frac54\,\s_\mu \s_\nu \C_P^{\mu\nu}   \;.
\end{eqnarray}
In the presence of interactions, \eq\eqref{eom567} implies that the Wigner-function is not on-shell. However, as in the spin-1/2 case, one can show that only the on-shell parts contribute to the Boltzmann equation, so that we can write it in the form of \eq\eqref{boltzonshell} with $f$ formally given by \eq\eqref{ff}. This will be shown in a forthcoming publication~\cite{forth}. The explicit form of the mass-shell correction of the spin-1 Wigner function does not play any role in the following discussion, since we will neglect off-shell effects in the conserved currents, as we did in Section \ref{freeprocsec}.

\subsection{Canonical currents}

Since we assume that $\mathcal{L}_\text{int}$ does not depend on derivatives of the fields, the canonical currents in the interacting case are formally still given by \eqs\eqref{S_C_Wigner}. Using the constraint equations \eqref{constr_coll} and relations \eqref{fkgfkf} we can write them in terms of the distribution function as
\begin{subequations}
\begin{eqnarray}
\TP{C}^{\mu\nu}&=& \int d\Gamma\, p^\mu p^\nu f +\mathcal{O}(\hbar^2)\;,\\
\SP{C}^{\lambda\mu\nu}&=&\int d\Gamma\, \left[p^\lambda \left(\Sigma_\ms^{\mu\nu}-\frac{\hbar}{6m^2} p^{[\mu} \partial^{\nu]}\right)+\frac12 p^{[\mu} \Sigma_\ms^{\nu]\lambda} +\frac{\hbar}{6}  P^{\lambda[\mu} \partial^{\nu]} \right] f\;. \label{S_can_1}
\end{eqnarray}
\end{subequations}
The canonical spin tensor for Proca fields is hence not formally equivalent to the one for Dirac fields in \eq\eqref{scan33}. In particular, as expected, it is not totally antisymmetric.

\subsection{Hilgevoord-Wouthuysen currents}
In order to obtain the HW pseudo-gauge transformations in the interacting case, we modify \eqs\eqref{HW_free_Proca} as
\begin{subequations}\label{HW_int_Proca}
\begin{eqnarray}
\hat{\Phi}^{\lambda,\mu\nu}_{HW}&=&  \hat{M}^{[\mu\nu]\lambda}-g^{\lambda[\mu} \hat{M}_\rho^{\;\;\nu]\rho}+\frac{\hbar^2}{m^2} g^{\lambda[\mu}\left( V^{\nu]}\partial\cdot \rho^\dagger+\mathrm{h.c.} \right)  \;,\\
\hat{Z}^{\mu\nu\lambda\rho}_{HW}&=&-\frac{1 }{2} \left(V^{\dagger[\mu} g^{\nu][\lambda} V^{\rho]}+\mathrm{h.c.}\right)\;.
\end{eqnarray}
\end{subequations}
In terms of the Wigner function, these pseudo-gauge potentials read
\begin{subequations}\label{HW_int_Proca_Wigner}
\begin{eqnarray}
\Phi_{HW}^{\lambda,\mu\nu}&=&\int \d^4 p \left(\frac{\hbar}{2}\partial_\rho \WP{S}^{\rho[\mu}g^{\nu]\lambda}+ip^{[\mu}\WP{A}^{\nu]\lambda}  \right)\; ,\\
Z_{HW}^{\mu\nu\lambda\rho}&=&\frac{1}{2}  \int \d^4 p \left(g^{\nu[\lambda}\WP{S}^{\rho]\mu}-g^{\mu[\lambda}\WP{S}^{\rho]\nu}\right)\;,
\end{eqnarray}
\end{subequations}
where we have made use of the constraint equations \eqref{boppconsint}. Note that the dependence of the pseudo-gauge potentials on the Wigner function in \eq\eqref{HW_int_Proca_Wigner} is identical to the noninteracting case. 
Equations \eqref{HW_int_Proca_Wigner} imply the following relations,
\begin{subequations}
\begin{eqnarray}
\hbar\, \partial_\rho Z_{HW}^{\mu\nu\lambda\rho}&=&\frac12\int \d^4 p \left(  \hbar\partial_\rho \WP{S}^{\rho[\mu}g^{\nu]\lambda} +\hbar\partial^{[\mu}W^{\nu]\lambda}_S \right)\; , \\
\Phi_{HW}^{\lambda,\mu\nu}+\Phi_{HW}^{\mu,\nu\lambda}+\Phi_{HW}^{\nu,\mu\lambda}&=&2\int \d^4 p\left( \frac{\hbar}{2}\partial_\rho \WP{S}^{\rho[\mu}g^{\lambda]\nu}+ip^{\nu}\WP{A}^{\lambda\mu} \right)\; ,
\end{eqnarray}
\end{subequations}
from which, after using the equations of motion, the HW energy-momentum tensor in the interacting case is obtained as
\begin{align}
\TP{HW}^{\mu\nu}={}& \int \d^4 p \left[\left(p^\mu p^\nu +\frac{\hbar^2}{4}\partial^\mu \partial^\nu \right) \mathrm{Tr}\,W_P -\left(p_\rho p^\nu-\frac{\hbar^2}{4}\partial_\rho \partial^\nu\right) \WP{S}^{\rho\mu} +\frac{i\hbar}{2} p^{(\nu}\partial_{\rho)}\WP{A}^{\rho\mu} \right] \nonumber \\
&+\hbar g^{\mu\nu}\left\langle: \left[ (\partial_\alpha V_\beta^\dagger)\partial^\alpha V^\beta -\frac{m^2}{\hbar^2}V^{\dagger\alpha}V_\alpha \right]:\right\rangle\label{T_spin1_int_HW_1}\;.
\end{align}
Making use of the constraint equations \eqref{constr2}, \eq\eqref{T_spin1_int_HW_1} becomes 
\begin{eqnarray}
\TP{HW}^{\mu\nu}
&=& \int \d^4 p \bigg[\left(p^\mu p^\nu +\frac{\hbar^2}{4}\partial^\mu \partial^\nu \right)\mathrm{Tr}\,W_P -p^\nu\frac{\hbar}{2m^2}\left( -p \C_A^{\mu} -p\D_S^{\mu} -2 p^\mu \D_E + \hbar\partial_\alpha P^{\alpha\mu}\C_P+\hbar\partial_\alpha E^{\mu\nu} \C_E\right)  \nonumber\\
&&-\,\frac{g^{\mu\nu}}{2}\left(p^2-m^2+\frac{\hbar^2}{4}\partial^2\right)\mathrm{Tr}\,W_P \bigg] +\mathcal{O}(\hbar^3) \nonumber\\
&=& \int \d^4 p \bigg[p^\mu p^\nu \left(3f_P+\frac{\hbar^2}{4m^2}P^{\alpha\beta} \partial_\alpha \partial_\beta f_P^{(0)}\right)+\frac{3\hbar^2}{4}\partial^\mu\partial^\nu f_P^{(0)}  +p^\nu\frac{\hbar^2}{2m^2}\left( p \C_A^{(1)\mu} +p\D_S^{(1)\mu} + \partial_\alpha P^{\alpha\mu}\C_P^{(0)}\right)  \nonumber\\
&&-\,\frac{g^{\mu\nu}}{2}\left(p^2-m^2+\frac{\hbar^2}{4}\partial^2\right)3f_P \bigg] +\mathcal{O}(\hbar^3) \;,
\end{eqnarray}
where we used $\C_E^{(0)}=\D_E^{(0)}=0$, see Sec. \ref{sec:pr_int}.
Up to first order in $\hbar$, the energy-momentum tensor is symmetric and formally equivalent to \eq\eqref{KleinGordontensors},
\begin{equation}
\TP{HW}^{\mu\nu}=\int d\Gamma\, p^\mu p^\nu f(x,p,\s)+\mathcal{O}(\hbar^2)\;.\label{T_HW_sym}
\end{equation}
Furthermore, the spin tensor up to first order is obtained by using \eq\eqref{81c} as
\begin{align}
    \SP{HW}^{\lambda,\mu\nu}&=2i \int \d^4 p\, p^\lambda \WP{A}^{\mu\nu} \nonumber\\
    &=\int d\Gamma\; p^\lambda \left( \Sigma_\s^{\mu\nu} -\frac{\hbar}{6m^2}{p^{[\mu}}\partial^{\nu]}   \right)f(x,p,\s)+\mathcal{O}(\hbar^2)\; . \label{sphw1fin}
\end{align}
Note that the HW spin tensor for Proca fields has the same structure as the one for Dirac fields in \eq\eqref{SpinHW}. The difference in the factors of the last terms of \eqs\eqref{SpinHW} and \eqref{sphw1fin}, respectively, is due to the different normalizations of the phase-space volume. After performing the integrations over $dS$ the  factor in both expressions will be 1/2.
At second order in $\hbar$, the energy-momentum tensor acquires an antisymmetric contribution due to interactions,
\begin{equation}
\TP{HW}^{[\mu\nu]}=  \hbar^2\int \d^4 p\, \frac{p^{[\nu}}{2m^2} \left( p\, \C_A^{(1)\mu]}+p\, \D_S^{(1)\mu]} +\partial^{\mu]} \C_P^{(0)}  \right)+\mathcal{O}(\hbar^3)\;,\label{T_HW_A}
\end{equation}
leading to the nonconservation of the spin tensor \eqref{sphw1fin}.
\subsection{Alternative Klein-Gordon currents}
In the interacting case, we modify the KG pseudo-gauge transformations in \eqs\eqref{pgtkgproca} according to
\begin{subequations}\label{KG_int_Proca}
\begin{eqnarray}
\hat{\Phi}_{KG}^{\lambda,\mu\nu}&=& \hat{M}^{[\mu\nu]\lambda}-g^{\lambda[\mu} \hat{M}_\rho^{\;\;\nu]\rho}+\frac{\hbar^2}{m^2}g^{\lambda[\mu}\left(V^{\nu]} \partial \cdot \rho^\dagger+\mathrm{h.c.}\right)-\frac{1}{2}g^{\lambda[\mu}\partial^{\nu]}V^{\dagger\beta} V_\beta \;,\\
\hat{Z}_{KG}^{\mu\nu\lambda\rho}&=&-\frac{1}{2}\left(V^{\dagger[\mu} g^{\nu][\lambda} V^{\rho]}+ \mathrm{h.c.}\right) - \frac{1}{4}\delta_\alpha^{[\nu} g^{\mu][\lambda} g^{\rho]\alpha} V^{\dagger\beta} V_\beta\;.
\end{eqnarray}
\end{subequations}
These pseudo-gauge potentials differ from \eqs \eqref{HW_int_Proca} by the addition of the last terms in each equation.

The KG energy-momentum tensor in the interacting case will consequently be given by
\begin{align}
\TP{KG}^{\mu\nu}={}& \int \d^4 p \left[p^\mu p^\nu \mathrm{Tr}\,W_P -\left(p_\rho p^\nu-\frac{\hbar^2}{4}\partial_\rho \partial^\nu\right) \WP{S}^{\rho\mu} +\frac{i\hbar}{2} p^{(\nu}\partial_{\rho)}\WP{A}^{\rho\mu} \right] \nonumber \\
&-\frac{\hbar g^{\mu\nu}}{2}\left\langle: \left[ V^\dagger_\alpha\left(\partial^2+\frac{m^2}{\hbar^2}\right)V^\alpha +\mathrm{h.c.} \right]:\right\rangle\label{T_spin1_int_KG_1}\;.
\end{align}
Using the constraint equations \eqref{constr2}, \eq\eqref{T_spin1_int_KG_1} becomes 
\begin{eqnarray}
\TP{KG}^{\mu\nu}
&=& \int \d^4 p \bigg[p^\mu p^\nu \left(3f_P+\frac{\hbar^2}{4m^2}P^{\alpha\beta} \partial_\alpha \partial_\beta f_P^{(0)}\right) +p^\nu\frac{\hbar^2}{2m^2}\left( p \C_A^{(1)\mu} +p\D_S^{(1)\mu} + \partial_\alpha P^{\alpha\mu}\C_P^{(0)}\right)   \nonumber \\
&&-\,\frac{g^{\mu\nu}}{2}\left(p^2-m^2-\frac{\hbar^2}{4}\partial^2\right)3f_P \bigg] +\mathcal{O}(\hbar^3) ,
\end{eqnarray}
which is manifestly symmetric up to order $\mathcal{O}(\hbar)$.
As in the free case, $\SP{KG}^{\lambda,\mu\nu}= \SP{HW}^{\lambda,\mu\nu}$  at any order in $\hbar$.
Consequently, the antisymmetric part of the KG energy-momentum tensor takes on the same form as in the HW pseudo-gauge \eqref{T_HW_A}.

\section{Equations of motion}
\label{eomsec}

Since the HW and KG energy-momentum and spin tensors for spin-1/2 and spin-1 particles are given by the same expressions, they formally follow the same equations of motion, although the explicit forms of the distribution functions and collision terms differ between the two cases~\cite{Weickgenannt:2021cuo,forth}. Using the
Boltzmann equation \eqref{boltzonshell}  we obtain the equations of motion presented in Ref.~\cite{Weickgenannt:2020aaf},
\begin{subequations}
\begin{align} \label{Ta}
 \partial_\mu T^{\mu\nu}_{{HW}} &  = \int d\Gamma\, p^\nu\,  {\mC}[f] = 0\;,\\
  \hbar\, \partial_\lambda S_{{HW}}^{\lambda,\mu\nu}
 & = \int d\Gamma\, \hbar \sigma \Sigma_\ms^{\mu\nu}\, {\mC}[f]
 = T_{{HW}}^{[\nu\mu]}\;, \label{total_cons}
\end{align}
\end{subequations}
where $\sigma=1/2$ or $\sigma=1$ for spin-1/2 and spin-1 particles, respectively.
As pointed out in Ref.~\cite{Weickgenannt:2020aaf}, the energy-momentum tensor is conserved as $p^\mu$ is a collisional invariant, while in general the spin tensor is not conserved due to the mutual conversion between spin and orbital angular momentum in nonlocal collisions. In the presence of nonlocal collisions, the HW spin tensor is conserved only in global equilibrium, when the process of aligning spin with vorticity has stopped and the collision term vanishes.
In global equilibrium the distribution function up to first order in $\hbar$ is given by
\begin{equation}\label{f_eq}
 f_{eq}(x,p,\ms)=\frac{1}{(2\pi\hbar)^3}\exp\left[-\beta (x)\cdot p
 +\frac\hbar2 \sigma \varpi_{\mu\nu}\Sigma_\ms^{\mu\nu}\right]\;,
\end{equation}
where $\beta^\mu\equiv u^\mu/T$, with $u^\mu$ being the fluid velocity and $T$ the temperature, and $\varpi_{\mu\nu}\equiv-(1/2)\partial_{[\mu}\beta_{\nu]}$ ~\cite{Florkowski:2017ruc,Becattini:2018duy,Florkowski:2018fap}. For a derivation of an exact solution for the Wigner function in global equilibrium see Ref.~\cite{Palermo:2021hlf}. Note that $\beta^\mu$ satisfies the Killing condition $\partial_{(\mu}\beta_{\nu)}=0$.
The equilibrium distribution function \eqref{f_eq} is obtained from the requirement that the collision term vanishes \cite{Weickgenannt:2020aaf,Weickgenannt:2021cuo}.
Inserting \eq\eqref{f_eq} into \eqs\eqref{SpinHW} or \eqref{sphw1fin} we obtain the expression for the HW spin tensor in equilibrium to leading order in $\hbar$,
\begin{equation} \label{equilibriumspintensor}
 S^{\lambda,\mu\nu}_{{HW},eq}  = \frac\hbar g \sigma \, n^{(0)} u^\lambda\varpi^{\mu\nu}\;,
\end{equation}
where $n^{(0)}\equiv g\int dP\, p\cdot u \, f_{eq}^{(0)}(x,p) $ is the zeroth-order particle density and $g\equiv \int dS$ is the number of internal degrees of freedom. The spin
tensor \eqref{equilibriumspintensor} has the same form as that used in the
formulation of relativistic spin hydrodynamics in Ref.~\cite{Florkowski:2017ruc}.

In contrast to the physical interpretation of \eq\eqref{total_cons}, which relates the divergence of the spin tensor directly to the nonconservation of $\Sigma^{\mu\nu}_\s$ in collisions and vanishes in global equilibrium, the equations of motion for the canonical spin tensor acquire additional terms. In particular, the canonical spin tensor is not conserved even in global equilibrium. Using \eq\eqref{f_eq} in \eq\eqref{scan332}, or \eq\eqref{f_eq} in \eq\eqref{S_can_1}, respectively, we obtain, cf.\ Ref.~\cite{Speranza:2020ilk}, 
\begin{equation}
\partial_\lambda S_{C,eq}^{\lambda,\mu\nu} = \frac{1}{(2\pi\hbar)^3} {\hbar}\sigma \int d P p^{[\mu} \varpi^{\nu]\lambda} p^\rho \varpi_{\lambda\rho}  e^{-\beta\cdot p} +\mathcal{O}(\hbar^2)\;,\label{div_scan_12}
\end{equation}
where it has been used that $\mathfrak{C}[f_{eq}]=0$.

\section{Including electromagnetic fields}
\label{emfsec}

So far, we have discussed the effects of a general collision term on the conserved currents without considering the interaction with gauge fields. In this section, we include electromagnetic fields and study their impact on the energy-momentum and spin tensors. In this case, a further complication arises since gauge invariance of the theory has to be guaranteed. The relativistic decomposition of orbital and spin angular momentum in the presence of gauge fields is a long-standing problem with consequences in different fields such as hadron and chiral physics, see, e.g., Refs.~\cite{Leader:2013jra,Fukushima:2020qta} for reviews. In the following, we will introduce a pseudo-gauge which combines a KG transformation for the matter parts of the currents with a Belinfante transformation for the electromagnetic parts. In this way, we obtain a gauge-invariant splitting of the total angular momentum with a vanishing gauge-field spin tensor.  For the sake of simplicity, we neglect particle collisions and treat  the electromagnetic fields as classical. Furthermore, we absorb the electromagnetic charge $e$ in the definition of the gauge potential. The currents and equations of motion derived in this section provide the starting point for the formulation of spin magnetohydrodynamics for Dirac and Proca particles.

\subsection{Dirac fields}

The gauge-invariant Wigner function for fermions interacting with the electromagnetic potential $\mathbb{A}^\mu(x)$ is defined as 
\cite{Heinz:1983nx,Vasak:1987um},
\begin{equation}
  \WD{\alpha\beta}(x,p)=\int \frac{d^4y}{(2\pi\hbar)^4} e^{-\frac i\hbar p\cdot y}
  \left\langle : \bar{\psi}_\beta\left(x_1\right)U(x_1,x_2)\psi_\alpha\left(x_2\right):\right\rangle \; , \label{wigwig4}
\end{equation}
where the gauge link 
\begin{equation}
 U(x_1,x_2)\equiv\exp\left[-\frac{i}{\hbar}y^\mu\int_{-1/2}^{1/2} dt\, \mathbb{A}_\mu\left(x+t y\right)\right] \label{gaugelink}
\end{equation}
ensures gauge invariance of the Wigner function.
The equations of motion read
\begin{equation}
    \left[\gamma\cdot\left(\ePi+\frac{i\hbar}{2}\Dd \right)-m\right]W_D=0\; ,
\end{equation}
with
\begin{equation}
\Dd^\mu\equiv \partial^\mu-j_0\left(\frac\hbar2\partial\cdot\partial_p\right)F^{\mu\nu}\partial_{p\nu}
\end{equation}
and
\begin{equation}
    \ePi^\mu\equiv p^\mu -\frac\hbar2j_1\left(\frac\hbar2\partial\cdot\partial_p\right)F^{\mu\nu}\partial_{p\nu} \;,
\end{equation}
where $j_0(x)\equiv \sin x/x$ and $j_1(x)\equiv (\sin x-x\cos x)/x^2$ are spherical Bessel functions and $F^{\mu\nu}\equiv \partial^{[\mu} \mathbb{A}^{\nu]}$ is the electromagnetic field-strength tensor.  The spacetime derivatives in the arguments of the spherical Bessel functions act only on the field-strength tensor, but not on the Wigner function.
The equations of motion for the Wigner-function components can be found in Refs.~\cite{Vasak:1987um,Weickgenannt:2019dks}. Here, we only display those which will be used in the following, namely 
\begin{subequations}
\begin{eqnarray}
\ePi_{\mu} \F-\frac12\hbar\Dd^\nu \Sc_{\nu\mu}-m\V_\mu&=&0\; , \label{Vem}\\
-\frac12\hbar\Dd_\mu \Pc+\frac12\epsilon_{\mu\beta\nu\sigma}\ePi^\beta \Sc^{\nu\sigma}+m\A_\mu&=&0\; ,\label{Aem}\\
\frac12\hbar(\Dd_\mu \V_\nu-\Dd_\nu \V_\mu)-\epsilon_{\mu\nu\alpha\beta}\ePi^{\alpha} \A^\beta-m\Sc_{\mu\nu}&=&0 \;, \label{Sem}\\
\frac12\hbar\Dd_\mu \F+\ePi^\nu \Sc_{\nu\mu}&=&0\; .\label{Bem}
\end{eqnarray}
\end{subequations}
The canonical energy-momentum and spin tensors of both matter and gauge fields were found to be~\cite{Weickgenannt:2019dks}
\begin{subequations}
\begin{align}
\TD{C}^{\mu\nu}&= \int d^4p\, \V^\mu \left(p^\nu + \mathbb{A}^\nu\right)-F^{\mu\lambda}\partial^\nu \mathbb{A}_\lambda+\frac14 g^{\mu\nu} F^{\alpha\beta} F_{\alpha\beta}\;,\\
\SD{C}^{\lambda,\mu\nu}&= -\frac12 \epsilon^{\lambda\mu\nu\rho} \int d^4p\, \A_\rho-\frac{1}{\hbar}F^{\lambda[\mu} \mathbb{A}^{\nu]}\;,
\end{align}
\end{subequations}
which are gauge-dependent quantities. In the following, we will perform a pseudo-gauge transformation which leads to a gauge-invariant splitting between spin and orbital angular momentum of the matter and gauge-field parts. This is achieved by generalizing the KG transformation for spin-1/2 particles which leads to formally the same pseudo-gauge potentials in terms of the Wigner function as in the free case. Furthermore, we use a Belinfante pseudo-gauge transformation for the gauge fields in order the obtain a gauge-invariant result~\cite{Leader:2013jra}. The pseudo-gauge potentials for such a transformation read
\begin{subequations}
\begin{align}
\Phi^{\lambda,\mu\nu}_{KG,B}&=\frac{1}{2m} \int d^4p\, p^{[\mu} \Sc^{\nu]\lambda} -\frac{1}{\hbar} F^{\lambda[\mu} \mathbb{A}^{\nu]}\;,\\
Z^{\mu\nu\lambda\rho}_{KG,B}&= \frac{1}{4m}\epsilon^{\mu\nu\lambda\rho} \int d^4p\, \Pc\;.
\end{align}
\end{subequations}
Therefore, the spin tensor is given by
\begin{align}
\SD{KG}^{\lambda,\mu\nu}=& \frac{1}{2m} \epsilon^{\lambda\mu\nu\rho} \int d^4p\, \left( \frac12 \epsilon_{\rho\alpha\beta\gamma} \ePi^\alpha \Sc^{\beta\gamma}-\frac{{\hbar}}{2}\nabla_\rho \Pc\right)-\frac{1}{2m} \int d^4p\, \Sc^{\lambda[\mu} p^{\nu]}+ \frac{{\hbar}}{4m}\epsilon^{\mu\nu\lambda\rho} \partial_\rho \int d^4p\, \Pc\n\\
=& \frac{1}{2m} \int d^4p\, p^\lambda \Sc^{\mu\nu}, \label{skgem}
\end{align}
where we made use of \eq\eqref{Aem} and assumed that boundary terms vanish. Moreover, we obtain the energy-momentum tensor
\begin{align}
\TD{KG}^{\mu\nu}=& \int d^4p\, \V^\mu \left(p^\nu + \mathbb{A}^\nu\right)- F^{\mu\lambda}\partial^\nu \mathbb{A}_\lambda+\frac14 g^{\mu\nu} F^{\alpha\beta} F_{\alpha\beta}+\frac{{\hbar}}{2} \partial_\lambda \bigg( \frac{1}{2m}\int d^4p\, \Sc^{\lambda[\mu} p^{\nu]}- \frac{1}{\hbar}F^{\lambda[\mu} \mathbb{A}^{\nu]}\n\\
&+\frac{1}{2m}\int d^4p\, \Sc^{\nu[\mu} p^{\lambda]}-\frac{1}{\hbar}F^{\nu[\mu} \mathbb{A}^{\lambda]}+\frac{1}{2m}\int d^4p\, \Sc^{\mu[\nu} p^{\lambda]}-\frac{1}{\hbar}F^{\mu[\nu} \mathbb{A}^{\lambda]}\bigg)\n\\
=& \frac{1}{m} \int d^4p\, \left(p^\mu p^\nu \F+\frac{{\hbar}}{2} F^\nu_{\ \lambda} \Sc^{\lambda\mu} \right)- F^{\mu\lambda} F^\nu_{\ \lambda}+\frac14 g^{\mu\nu} F^{\alpha\beta} F_{\alpha\beta}\; \label{tkgem}
\end{align}
with the antisymmetric part
\begin{equation}
\TD{KG}^{[\mu\nu]}= \frac{{\hbar}}{2m} \int d^4p\, \Sc^{\lambda[\mu} F^{\nu]}_{\ \ \lambda}.
\end{equation}
When deriving \eq\eqref{tkgem}, we inserted \eq\eqref{Vem} and the Maxwell equation $\partial_\mu F^{\mu\nu}=J^\nu$,
where
\begin{equation}
 J_\mu\equiv  \int d^4p\, \V_\mu\;  \label{charcurrr}
\end{equation}
is the charge current, and again made use of the assumption of vanishing boundary terms. We see that both the energy-momentum and spin tensors are gauge invariant. 

The above currents are now separated into fluid and electromagnetic parts according to
\begin{align}
\TD{f}^{\mu\nu}&= \frac{1}{m} \int d^4p\, \left(p^\mu p^\nu \F+\frac{{\hbar}}{2} F^\nu_{\ \lambda} \Sc^{\lambda\mu} \right)\; ,\n\\
T^{\mu\nu}_{em}&= - F^{\mu\lambda} F^\nu_{\ \lambda}+\frac14 g^{\mu\nu} F^{\alpha\beta} F_{\alpha\beta}\; ,\n\\
\SD{f}^{\lambda,\mu\nu}&=  \frac{1}{2m} \int d^4p\, p^\lambda \Sc^{\mu\nu}\; ,\n\\
S^{\lambda,\mu\nu}_{em}&= 0\; .
\end{align}
In this case, the spin tensor for the electromagnetic fields vanishes and only fermionic spin degrees of freedom are treated as dynamical, while the electromagnetic ones are absorbed into the orbital angular momentum from the energy-momentum tensor.
We find the following equation of motion for the fluid energy-momentum tensor,
\begin{align}
\partial_\mu \TD{f}^{\mu\nu} =& \frac1m \int d^4p\, p^\mu p^\nu F_{\mu\lambda}\partial_p^\lambda \F +\frac{\hbar}{6m} \int d^4p\, p^\mu p^\nu (\partial^\alpha F^{\rho\lambda})\partial_{p\lambda}\partial_{p\alpha} \Sc_{\rho\mu}\n\\
&+ \frac{\hbar}{2m} (\partial_\mu F^\nu_{\  \lambda}) \int d^4p\, S^{\lambda\mu}+\frac{\hbar}{2m} \int d^4p\, F^\nu_{\ \lambda} \partial_\mu \Sc^{\lambda\mu}\n\\
=& -F^{\mu\nu} J_\mu, \label{teomem}
\end{align}
where we used \eqs\eqref{Bem}, \eqref{Vem}, and the Maxwell relation $\partial^{\mu} F^{\nu\lambda}+\partial^\nu F^{\lambda\mu}+\partial^\lambda F^{\mu\nu}=0$.
Since
\begin{equation}
\partial_\mu T^{\mu\nu}_{em}= F^{\mu\nu} J_\mu=-\partial_\mu \TD{f}^{\mu\nu},
\end{equation}
the total energy-momentum tensor is conserved. On the other hand, the spin tensor is not conserved but follows the equations of motion
\begin{equation}
\hbar\, \partial_\lambda  \SD{f}^{\lambda,\mu\nu}=-\frac{{\hbar}}{2m} \int d^4p\, \Sc^{\lambda[\mu} F^{\nu]}_{\ \ \lambda}=\TD{f}^{[\nu\mu]}. \label{seomem}
\end{equation}

We remark that the results of this section are similar to those
of Ref.~\cite{Israel:1978up} for fluids with polarization when identifying $\Sc^{\mu\nu}$ with the dipole moment~\cite{Weickgenannt:2019dks}, although the former are exact in $\hbar$ and the latter purely classical. In particular, as can be seen from  \eq\eqref{seomem}, the equations of motion for $s^{\mu\nu}_{HW}$, defined through
\begin{equation}
   \int d\Sigma_\lambda\, \SD{HW}^{\lambda,\mu\nu}= \int d^4p\, s_{D,HW}^{\mu\nu},
\end{equation}
where $d\Sigma_\lambda$ denotes the integration over a spacelike hypersurface, are the Matthison-Papapetrou-Dixon (MPD) equations~\cite{Bailey:1975fe,Israel:1978up}
\begin{align}
    m\frac{d}{d\tau} s^{\mu\nu}_{D,HW}=& \frac{1}{2m} \int d\Sigma_{\lambda} p^\lambda p^\rho \partial_\rho \Sc^{\mu\nu}\n\\
    =& -\frac{1}{2m} \int d\Sigma_{\lambda}p^\lambda \Sc^{\rho[\mu} F^{\nu]}_{\ \ \rho}\n\\
    =& -s_{D,HW}^{\rho[\mu} F^{\nu]}_{\ \ \rho}\; ,
\end{align}
where $\tau\equiv x\cdot p/m$ is the proper time.

\subsection{Proca fields}

In order to describe Proca fields interacting with electromagnetic fields we use a Lagrangian of the form~\cite{Corben.1940}
\begin{equation}
   \mathcal{L}_{P,em} = \hbar \left(-\frac{1}{2}V^{\dagger\mu\nu} V_{\mu\nu} +\frac{m^2 }{\hbar^2} V^{\dagger\mu} V_\mu \right)-\frac{1}{4\ }F^{\mu\nu} F_{\mu\nu} 
-  i  F_{\mu\nu} V^\mu V^{\dagger\nu}\; ,
\end{equation}
where in the presence of gauge fields
\begin{equation}
    V^{\mu\nu}\equiv \left(\partial^{[\mu} +\frac{i}{\hbar}\bA^{[\mu} \right) V^{\nu]}
\end{equation}
is defined with a covariant instead of a partial derivative.
The Wigner function in this case is given by \eq\eqref{Wigner_function} supplemented with a gauge link $U(x_1,x_2)$ which is identical to \eq\eqref{gaugelink},
\begin{equation}
    W^{\mu\nu}_P\equiv-\frac{2}{\hbar(2\pi\hbar)^4} \int \d^4 v\, e^{-ip\cdot v/\hbar} \avg{V^{\dagger\mu}\left(x_1\right) U(x_1,x_2)V^\nu\left(x_2\right)}\;.
\end{equation}
The detailed derivation of the equations of motion in this case will be presented in a future work \cite{forth}.
The canonical energy-momentum tensor reads
\begin{subequations}
\begin{eqnarray}
\T_{P,C}^{\mu\nu}
&=&-\hbar  \left( V^{\mu\rho}\partial^{\nu} V^\dagger_\rho +V^{\dagger\mu\rho} \partial^{\nu} V_\rho\right) - F^{\mu\rho}\partial^\nu \bA_\rho -i  V^{[\mu} V^{\dagger\rho]}\partial^\nu \bA_\rho-g^{\mu\nu} \mathcal{L}_{P,em}\; ,\\
\TP{C}^{\mu\nu}&=&  \int \d^4 p \left[ \left(p^\mu p^\nu +\frac{\hbar^2}{4}\partial^\mu \partial^\nu\right)\mathrm{Tr}\,W_P -\left(p^\nu p_\rho +\frac{\hbar^2}{4}\partial^\nu \partial_\rho \right)\WP{S}^{\rho\mu}-\frac{i\hbar}{2} \left(p^{[\nu} \partial_{\rho]}-F^\nu_{\ \rho}\right) \WP{A}^{\rho\mu} \right]+j^\mu \bA^\nu\n\\
&&+i\hbar \int d^4p\, \WP{A}^{\rho\mu}\, \partial^\nu \bA_\rho- F^{\mu\rho}\partial^\nu \bA_\rho-g^{\mu\nu}\langle: \mathcal{L}_{P,em}:\rangle\; ,
\end{eqnarray}
\end{subequations}
where we dropped boundary terms and defined
\begin{equation}
    j^\mu\equiv \int d^4p\, \left( p^\mu \mathrm{Tr}\,W_P -p_\alpha \WP{S}^{\alpha\mu}-\frac{i\hbar}{2}\partial_\alpha \WP{A}^{\alpha\mu}\right)\;. \label{charprocem}
\end{equation}
 Furthermore, the spin tensor is given by
\begin{subequations}
\begin{eqnarray}
 \Sp_{P,C}^{\lambda,\mu\nu}&=&- \left(V^{\lambda[\mu}V^{\dagger\nu]}+V^{\dagger\lambda[\mu}V^{\nu]}   \right) - \frac{1}{\hbar} F^{\lambda[\mu}\bA^{\nu]} -i\left(   V^\lambda V^{\dagger[\mu}-V^{\dagger\lambda} V^{[\mu}\right)\bA^{\nu]}\; ,\\
S_{P,C}^{\lambda,\mu\nu} &=& - i \int \d^4 p \left[ 2p^\lambda W^{\nu\mu}_A-p^{[\mu}W^{\nu]\lambda}_A+\frac{i\hbar}{2}\partial^{[\mu}W^{\nu]\lambda}_S    \right]+ i\bA^{[\nu}\int d^4p\, \WP{A}^{\mu]\lambda}- \frac{1}{\hbar}F^{\lambda[\mu}\bA^{\nu]}\; .
\end{eqnarray}
\end{subequations}
Also here the canonical currents are not gauge invariant.

Now we perform a suitable pseudo-gauge transformation to obtain Klein-Gordon currents in the interacting case. Analogously to the previous discussion, we perform a Belinfante transformation for the gauge-field part and hence use the following pseudo-gauge potentials,
\begin{subequations}
\begin{eqnarray}
\Phi^{\lambda,\mu\nu}_{KG,B}&=&\int \d^4 p\,\bigg(\frac\hbar2\partial_\rho \WP{S}^{\rho[\mu}g^{\nu]\lambda}+i p^{[\mu} \WP{A}^{\nu]\lambda}+ \frac{\hbar}{4}  \partial^{[\nu}g^{\mu]\lambda} W^\beta_{P\;\beta} \bigg)- \frac{1}{\hbar} F^{\lambda[\mu} \bA^{\nu]} -i \int \d^4 p\, \WP{A}^{\lambda[\mu} \bA^{\nu]}\; ,\\
Z^{\mu\nu\lambda\rho}_{KG,B}&=&\frac12  \int \d^4 p\, \left( g^{\nu[\lambda} \WP{S}^{\rho]\mu} -g^{\mu[\lambda} \WP{S}^{\rho]\nu} +\frac{1}{2}\delta_\alpha^{[\nu} g^{\mu][\lambda} g^{\rho]\alpha} \mathrm{Tr}\,W_P    \right)\; .
\end{eqnarray}
\end{subequations}
Employing
\begin{eqnarray}
 \hbar\, \partial_\rho Z^{\mu\nu\lambda\rho}_{KG,B}&=& \int \d^4 p\, \bigg(  \frac{\hbar}{2} \partial_\rho \WP{S}^{\rho[\mu}g^{\nu]\lambda}+\frac{\hbar}{2} \partial^{[\mu} \WP{S}^{\nu]\lambda}+\frac{\hbar}{4}  \partial^{[\nu}g^{\mu]\lambda} \mathrm{Tr}\,W_P   \bigg)\;,
\end{eqnarray}
we find for the KG spin tensor
\begin{equation}
\SP{KG}^{\lambda,\mu\nu}=2i\int \d^4 p\, p^\lambda \WP{A}^{\mu\nu}\; ,
\end{equation}
coinciding with our earlier results.
In order to obtain the energy-momentum tensor, we compute
\begin{eqnarray}
\frac\hbar2\left(\Phi^{\lambda,\mu\nu}_{KG,B}+\Phi^{\mu,\nu\lambda}_{KG,B}+\Phi^{\nu,\mu\lambda}_{KG,B}\right)&=&\int \d^4 p \left(\frac{\hbar^2}{2} \partial_\rho \WP{S}^{\rho[\mu} g^{\lambda]\nu}+i\hbar p^{\nu}W^{\lambda\mu}_A+ \frac{\hbar^2}{4} \partial^{[\lambda}g^{\mu]\nu} \mathrm{Tr}\,W_P \right)\nonumber\\
&&- F^{\lambda\mu}\bA^\nu-i\hbar \int \d^4 p\, \WP{A}^{\lambda\mu} \bA^\nu\;.
\end{eqnarray}
Considering the derivative of the last line of the above equation, we find
\begin{equation}
\partial_\lambda\left(- F^{\lambda\mu}\bA^\nu-i\hbar \int \d^4 p\, \WP{A}^{\lambda\mu} \bA^\nu\right)=-j^\mu \bA^\nu  -F^{\lambda\mu}\partial_\lambda \bA^\nu -i\hbar \int \d^4 p\,  \WP{A}^{\lambda\mu} \partial_\lambda \bA^\nu\;,
\end{equation}
where we used Maxwell's equations
\begin{equation}
    \partial_\lambda F^{\lambda\mu}=j^\mu+i\hbar\, \partial_\lambda \int \d^4 p\, \WP{A}^{\mu\lambda}\; ,
\end{equation}
see Ref.~\cite{forth} for details. Putting everything together, we obtain the KG energy-momentum tensor
\begin{eqnarray}
 \TP{KG}^{\mu\nu}&=& \int d^4p\, \bigg[ p^\mu p^\nu \mathrm{Tr}\,W_P -\left( p^\nu p_\rho -\frac{\hbar^2}{4}\partial^\nu\partial_\rho\right) \WP{S}^{\rho\mu}+\frac {i\hbar}{2}\left( p^{(\nu}\partial_{\rho)}+F^\nu_{\ \rho}\right) \WP{A}^{\rho\mu}\bigg]+F^{\mu\lambda}F_\lambda^{\;\;\nu}\n\\
 &&+i\hbar \int \d^4 p\,  \WP{A}^{\mu\lambda} F_\lambda^{\;\;\nu}-g^{\mu\nu}\left( \langle: \mathcal{L}_{P,em}:\rangle-\frac{\hbar^2}{4}\partial^2\int d^4p\, \mathrm{Tr}\,W_P \right)\n\\
 &=& \int d^4p\, \bigg[ p^\mu p^\nu \mathrm{Tr}\,W_P -\frac{\hbar^2}{m^2}p^\nu(\partial^\gamma F_{\gamma\delta}) \WP{S}^{\delta\mu}-\frac{i\hbar^3}{m^2}(\partial^\gamma F_{\gamma\delta})\partial^\nu \WP{A}^{\delta\mu}\bigg]+F^{\mu\lambda}F_\lambda^{\;\;\nu}\n\\
 &&+2i\hbar \int \d^4 p\,  \WP{A}^{\mu\lambda} F_\lambda^{\;\;\nu}-g^{\mu\nu}\left( \langle: \mathcal{L}_{P,em}:\rangle-\frac{\hbar^2}{4}\partial^2\int d^4p\, \mathrm{Tr}\,W_P \right)\;,
\end{eqnarray}
where in the last step we inserted the constraint equations~\cite{forth}
\begin{subequations}\label{Kstar_W_SA}
\begin{eqnarray}
\Pi_\alpha \WP{S}^{\mu\alpha}+\frac{i\hbar}{2}\nabla_\alpha \WP{A}^{\mu\alpha}
&=& \frac{\hbar^2}{m^2} \left[ \cos\left(\frac\hbar2\partial\cdot\partial_p\right)(\partial^\gamma F_{\gamma\delta}) \WP{S}^{\delta\mu}+i\sin\left(\frac\hbar2\partial\cdot\partial_p\right)(\partial^\gamma F_{\gamma\delta}) \WP{A}^{\delta\mu} \right]\;, \\
\frac{i\hbar}{2} \nabla_\alpha \WP{S}^{\mu\alpha}+\Pi_\alpha \WP{A}^{\mu\alpha}
&=& \frac{\hbar^2}{m^2} \left[ i\sin\left(\frac\hbar2\partial\cdot\partial_p\right)(\partial^\gamma F_{\gamma\delta}) \WP{S}^{\delta\mu}+\cos\left(\frac\hbar2\partial\cdot\partial_p\right)(\partial^\gamma F_{\gamma\delta}) \WP{A}^{\delta\mu} \right] \;,
\end{eqnarray}
\end{subequations}
integrated by parts and neglected boundary terms. The form of the KG currents resembles the one obtained in the previous section, where for Proca fields the antisymmetric part of the Wigner function plays the role of the dipole moment.
While the equations of motion for the energy-momentum tensor are equivalent to those for spin 1/2, the equation of motion for the spin-1 tensor contains additional terms at quantum level, and the MPD equations are recovered only at the leading order,
\begin{eqnarray}
     \hbar\,  \partial_\lambda \SP{f}^{\lambda,\mu\nu}&=& 2i\hbar F_\lambda^{\;\;[\mu} \int \d^4 p\,  \WP{A}^{\nu]\lambda}+ \int d^4p\, \bigg[ \frac{\hbar^2}{m^2}(\partial^\gamma F_{\gamma\delta}) \WP{S}^{\delta[\mu}p^{\nu]}-\frac{i\hbar^3}{m^2}(\partial^\gamma F_{\gamma\delta})\partial^{[\nu} \WP{A}^{\mu]\delta}\bigg]   \; .
\end{eqnarray}
This result is to be expected, as spin-1 particles not only possess an intrinsic magnetic dipole moment, but also an electric quadrupole moment, which influences the spin dynamics at higher order in $\hbar$.

\section{Conclusions}

In this paper, we provided the explicit expressions of the pseudo-gauge transformations for the HW, GLW, and KG currents for interacting Dirac and Proca fields. For both spin-1/2 and spin-1 particles the spin tensor in such pseudo-gauges is conserved for free fields or for local interactions, but in general it is not in the presence of nonlocal collisions. Under a suitable definition of the enlarged phase space, the form of these currents for spin-1/2 and spin-1 particles differs only by degeneracy or spin-magnitude factors.  Considering electromagnetic interactions, we found a gauge-invariant splitting of the total angular momentum by performing a KG pseudo-gauge transformation for the matter fields and a Belinfante pseudo-gauge transformation for the gauge fields. The equations of motion of the spin tensor can then be related to the MPD equations. The energy-momentum and spin tensors for interacting systems derived in this work have a natural physical interpretation and provide the starting point to formulate spin (magneto-)hydrodynamics for Dirac ~\cite{Weickgenannt:2020aaf,Weickgenannt:2021cuo,Weickgenannt:2022zxs} and Proca fields. 

\section*{Acknowledgments} 

The authors thank D.\ H.\ Rischke and G.\ Torrieri for enlightening discussions.
The work of D.W.\ and N.W.\ is supported by the
Deutsche Forschungsgemeinschaft (DFG, German Research Foundation)
through the Collaborative Research Center CRC-TR 211 ``Strong-interaction matter
under extreme conditions'' -- project number 315477589 - TRR 211 and by the State of Hesse within the Research
Cluster ELEMENTS (Project ID 500/10.006). D.W. acknowledges support by the Studienstiftung des deutschen
Volkes (German Academic Scholarship Foundation).

\bibliography{biblio_paper_long}{}

\end{document}